\documentstyle [12pt] {article}

\newcommand{\gtwid}{\mathrel{\raise.3ex\hbox{$>$\kern-.75em\lower1ex
\hbox{$\sim$}}}}
\newcommand{\ltwid}{\mathrel{\raise.3ex\hbox{$<$\kern-.75em\lower1ex
\hbox{$\sim$}}}}
\newcommand{\beq}{\begin{equation}}
\newcommand{\eeq}{\end{equation}}
\newcommand{\beqs}{\begin{eqnarray}}
\newcommand{\eeqs}{\end{eqnarray}}

\catcode`@=11
\@addtoreset{equation}{section}
\@addtoreset{equation}{subsection}
\def\theequation{\ifnum\value{section}=0 \arabic{equation}\ignorespaces
\else \ifnum\value{section}=-1 A.\arabic{equation}\ignorespaces 
\else \ifnum\value{subsection}=0 \thesection.\arabic{equation}\ignorespaces
\else \thesection.\arabic{subsection}.\arabic{equation}\ignorespaces
                           \fi
                      \fi
                 \fi}
\catcode`@=12

\begin{document}

\def\thefootnote{\fnsymbol{footnote}}
\baselineskip 6.0mm

\begin{flushright}
\begin{tabular}{l}
ITP-SB-94-57    \\
December, 1994
\end{tabular}
\end{flushright}

\vspace{8mm}
\begin{center}
{\Large \bf Complex-Temperature Properties of }  

\vspace{3mm}

{\Large \bf the 2D Ising Model with $\beta H = \pm i \pi/2$} 

\vspace{16mm}

\setcounter{footnote}{0}
Victor Matveev\footnote{email: vmatveev@max.physics.sunysb.edu}
\setcounter{footnote}{6}
and Robert Shrock\footnote{email: shrock@max.physics.sunysb.edu}

\vspace{6mm}
Institute for Theoretical Physics  \\
State University of New York       \\
Stony Brook, N. Y. 11794-3840  \\

\vspace{16mm}

{\bf Abstract}
\end{center}
     We study the complex-temperature properties of a rare example of a
statistical mechanical model which is exactly solvable in an external 
symmetry-breaking field, namely, the Ising model on the square lattice 
with $\beta H = \pm i \pi/2$.  This model was solved by Lee and Yang 
\cite{ly}.  We first determine the complex-temperature phases and their 
boundaries. From a low-temperature, high-field series expansion of the 
partition function, we extract the low-temperature series 
for the susceptibility $\chi$ to $O(u^{23})$, where $u=e^{-4K}$.  
Analysing this series, we conclude that $\chi$ has divergent singularities (i) 
at $u=u_e=-(3-2^{3/2})$ with exponent $\gamma_e'=5/4$, (ii) at $u=1$, with
exponent $\gamma_1'=5/2$, and (iii) at $u=u_s=-1$, with exponent
$\gamma_s'=1$. We also extract a shorter series for the staggered 
susceptibility and investigate its singularities.  Using the exact result of 
Lee and Yang for the free energy, we calculate the specific heat and determine
its complex-temperature singularities.  We also carry this out for the uniform
and staggered magnetisation. 

\vspace{16mm}

\pagestyle{empty}
\newpage

\pagestyle{plain}
\pagenumbering{arabic}
\renewcommand{\thefootnote}{\arabic{footnote}}
\setcounter{footnote}{0}

\section{Introduction}
\label{intro}

    The Ising model has long served as a prototype of a statistical mechanical
system which undergoes a phase transition with associated spontaneous symmetry
breaking and long range order.  In the absence of an external magnetic field
$H$, the free energy of the $d=2$ (spin $1/2$) Ising model was first 
calculated by Onsager \cite{ons}, and the expression for the spontaneous 
magnetisation first calculated by Yang \cite{yang}, both 
for the square lattice.  The model has never been solved in an arbitrary 
external magnetic field.  However, in one of their classic papers, 
Lee and Yang \cite{ly} did succeed in solving exactly for the free energy and 
giving an exact expression for the magnetisation of the Ising model on 
the square lattice for a particular manifold of values of $H$ depending on 
the temperature $T$, given by 
\beq
H = \frac{i \pi k_B T}{2}
\label{hvalue}
\eeq
Although this is not a physical set of values, owing to the imaginary
value of $H$ and the resultant non-hermiticity of the Hamiltonian, this 
model is nevertheless of considerable interest as a rare example of a
statistical mechanical model for which one has an exact solution in the
presence of a symmetry-breaking field.  Further work on the 
derivation of the Lee-Yang solution was reported in Refs. 
\cite{gb,mw,merlini,lw}.

   In the present paper, we shall investigate this model in a wider context, 
generalising the temperature to complex values.  There are several reasons 
for studying the properties of statistical mechanical systems with the 
temperature variable generalised to take on complex values.  First, one can 
understand more deeply the behaviour of various 
thermodynamic quantities by seeing how they behave as analytic functions of
complex temperature.  Second, one can see how the physical phases of a given
model generalise to regions in appropriate complex-temperature variables.
Third, a knowledge of the complex-temperature singularities of quantities
which have not been calculated exactly helps in the search for exact,
closed-form expressions for these quantities.  This applies, in particular,
to the susceptibility of the present model, which, like that of the
zero-field Ising model, has never been calculated. 

\section{Generalities and Complex-Temperature Phases}
\label{general}

    In this section we shall work out the complex-temperature phases and their
boundaries.  We begin with some definitions and notation.  Recall that 
the Ising model is defined by the partition function 
$Z = \sum_{\{\sigma_n\}} e^{-\beta {\cal H}}$ with the Hamiltonian
\beq
{\cal H} = -J \sum_{<n n'>} \sigma_n \sigma_{n'} - H \sum_n \sigma_n
\label{h}
\eeq
where $\sigma_n = \pm 1$ are the $Z_2$ spin variables on each site $n$ of the
lattice, $\beta = (k_BT)^{-1}$, $J$ is the exchange constant, $<nn'>$ denote 
nearest-neighbor pairs, and the units are defined such that the magnetic 
moment which would multiply the $H\sum_n \sigma_n$ is unity.  
We shall concentrate here on the square (sq) lattice.  We use the 
standard notation $K = \beta J$, $h = \beta H$, $v = \tanh K$, $z = e^{-2K}$, 
$u = z^2 = e^{-4K}$, $w=1/u$, and $\mu = e^{-2h}$.  
Note that $v$ and $z$ are related 
by the bilinear conformal transformation
\beq
z = \frac{1-v}{1+v}
\label{bilinear}
\eeq
It will also be useful to introduce two elliptic moduli.  The first is 
\beq
\kappa = \frac{1}{C^2} = \frac{4u}{(1+u)^2}
\label{kappa}
\eeq
which occurs in elliptic integrals in the exact expressions for the internal 
energy and specific heat, where we use the abbreviations 
\beq
C \equiv \cosh(2K)
\label{cosh2k}
\eeq
\beq
S \equiv \sinh(2K)
\label{sinh2k}
\eeq
We record the symmetry 
\beq
u \to 1/u \qquad \Rightarrow \quad \kappa \to \kappa 
\label{uksym}
\eeq
The second elliptic modulus, 
\beq
k_< = \frac{i}{S(S^2+2)^{1/2}} = \frac{4iu}{(1-u)(1+6u+u^2)^{1/2}}
\label{kl}
\eeq
occurs in a natural way in the magnetisation.\footnote{Note that these differ 
from the respective elliptic moduli $\kappa_0$ and $k_{<,0}$ which occur in the
internal energy, specific heat, and spontaneous magnetisation for the Ising
model on the square lattice at $h=0$.} 

The reduced free energy per site is
$f = -\beta F = \lim_{N_s \to \infty} N_s^{-1} \ln Z$
(where $N_s$ denotes the number of sites on the lattice).  In addition to the
susceptibility itself, it will also be convenient to refer to the reduced 
susceptibility $\bar\chi = \beta^{-1}\chi$. 

   We begin by discussing the phase boundaries of the model as a function of
complex temperature, i.e. the locus of points across which the free energy is
non-analytic.  As noted in Ref. \cite{ms}, there is an infinite periodicity 
in complex $K$ under the shift $K \to K + n i \pi$, where $n$ is an integer,
and, for lattices 
with even coordination number $q$, also the shift  $K \to (2n+1)i\pi/2$, as 
a consequence of the fact that the spin-spin interaction
$\sigma_i\sigma_j$ in ${\cal H}$ is an integer.  In particular, there is an
infinite repetition of phases as functions of complex $K$; these repeated
phases are reduced to a single set by using the variables $v$, $z$, or $u$ (or
variables based on these).

   We also note an elementary symmetry involving $h$ for the (spin 1/2) Ising
model on a general lattice $\Lambda$.  The low-temperature, high-field 
expansion of $Z$ has the form 
\beq
Z = e^{(q/2)N_s K}e^{N_s h}Z_r
\label{zexpansion}
\eeq
where
\beq
Z_r = 1 + \sum_{n,m} a_{n,m}^{(\Lambda)} z^{n} \mu^m
\label{zr}
\eeq
where the only property of $Z_r$ that we need is the fact that it is a 
polynomial in $z$ and $\mu$.  In 
(\ref{zexpansion}), we assume periodic boundary conditions, but
for the free energy, in the thermodynamic limit, this is not essential.  Also,
parenthetically, we note that for a lattice with even $q$, only even powers of
$z$ occur in $Z_r$, i.e. $Z_r$ is a series in $u$, but we shall not need this 
fact here.  Now
\beq
h \to h + n i \pi \quad \Rightarrow \quad \mu \to \mu
\label{hshift}
\eeq
where $n$ is an integer. 
Hence, under such a shift, the only change in $Z$ is in the prefactor,
$e^{N_sh}$.  Equivalently, in the corresponding (reduced) free energy
\beq
f = (q/2)K + h + \lim_{N_s \to \infty} N_s^{-1}(1 + \sum_{n,m} 
a_{nm}^{(\Lambda)} z^{n} \mu^m )
\label{fgeneral}
\eeq
the only change is in the second term, $h$.  Therefore, aside from this term, 
one may, and we shall, restrict to the range 
\beq
-\frac{i \pi}{2} < Im(h) \le \frac{i\pi}{2}
\label{hrange}
\eeq
without loss of generality.  In the present context, we shall consider just the
value $h=i\pi/2$; our results will apply in the same way to $h=-i\pi/2$. 

   It is useful to review the connection between the square-lattice Ising
model with $h=i\pi/2$ and the Ising model on
the square lattice in zero field \cite{merlini,lw}.  This is done by first 
considering the Ising model on the checkerboard (also called generalised 
square) lattice, defined by assigning different couplings $K_j$, $j=1,...,4$, 
to the bonds of the square lattice, as shown in Fig. 1.  Again, for discussions
of the partition function, we assume periodic boundary conditions. 
The free energy \cite{utiyama} and
spontaneous magnetisation \cite{sn,baxter} are known for the zero-field 
checkerboard Ising model.  Now recall the identity 
\beq
e^{h\sigma} = \cosh h + \sigma \sinh h
\label{identity}
\eeq
for $\sigma=\pm 1$.  For $h=i\pi/2$, this reduces to 
$e^{h\sigma_n}=i\sigma_n$, and hence
$\exp(h\sum_n \sigma_n) = \prod_{n} (i \sigma_n)$.   
Next, consider a dimer site covering of the checkerboard lattice, 
where by site covering, we mean that each site is the member of one (and 
only one) dimer.  As is clear from Fig. 1, a simple covering of this sort is 
provided by each of the bonds of a single type, say those with the $K_4$ 
couplings.  We may thus associate pairs of the $\sigma_n$'s in the above
product with the dimers of this covering.  To do this, we separate one of the
two factors of $i$ for such pair and place it in front of $Z$.  One then has 
\beq
Z_{ch} = i^{N_s/2}\sum_{ \{\sigma_n \} } 
\Bigl ( \prod_{<rs>}(i\sigma_r \sigma_s) 
\Bigr ) \exp(\sum_{<n,n'>}\sigma_n K_{n n'} \sigma_{n'})
\label{zfactor}
\eeq
where $ch$ denotes checkerboard and 
$K_{n n'}$ refers to the appropriate $K_j$, $j=1,2,3,4$ depending on which bond
connects the sites $n$ and $n'$ (c.f. Fig. 1). 
Then one can use the identity (\ref{identity}) again to write 
\beq
Z = i^{N_s/2}\sum_{ \{\sigma_n \} } \exp( \sum_{<n n'>} \sigma_n
K'_{n,n'}\sigma_{n'}) 
\label{zrelationa}
\eeq
that is, 
\beq
Z(\{K_i \};h=i \pi/2)_{ch} = (i)^{N_s/2}Z(\{K'_i\};h=0)_{ch}
\label{zrelation}
\eeq
where $K'_i = K_i$, $i=1,2,3$, and 
\beq
K'_4 =K_4 + \frac{i\pi}{2}
\label{k4p}
\eeq 
Hence, for the free energy, 
\beq
f(\{K_i \};h=i \pi/2)_{ch} = \frac{i\pi}{4} + f(\{K'_i\};h=0)_{ch}
\label{frelation}
\eeq

Then, setting $K_i=K$, $i=1,2,3,4$, one can obtain the Lee-Yang result for
$f(K,h=i\pi/2)$ \cite{ly} 
from the (analytic continuation of the) free energy for the 
zero-field checkerboard lattice \cite{utiyama}.  The same method works for the
magnetisation and yields the relation 
\beq
M(\{K_i\},h=i\pi/2)_{ch} = M(\{K'_i\},h=0)_{ch}
\label{magrel}
\eeq
and for the $m$-point correlation functions, which satisfy 
\beq
<\sigma_{n_1} \cdot \cdot \cdot \sigma_{n_m}>(\{ K_i \}, h=i\pi/2)_{ch} = \ 
<\sigma_{n_1} \cdot \cdot \cdot \sigma_{n_m}>(\{ K'_i \}, h=0)_{ch}
\label{corfunrel}
\eeq
Further, it follows from the special case of (\ref{corfunrel}) for 2-spin
correlation functions, together with the expression for the susceptibility 
as a sum over the connected (conn.) 2-spin correlation functions, 
\beq
\bar\chi = \sum_{\bf r} <\sigma_{\bf 0}\sigma_{\bf r}>_{conn.}
\label{chisum}
\eeq
(where $<\sigma_{\bf 0}\sigma_{\bf r}>_{conn.} \equiv 
<\sigma_{\bf 0}\sigma_{\bf r}> \ - M^2$) that
\beq
\bar\chi(\{K_i \},h=i\pi/2)_{ch} = \bar\chi(\{K'_i \},h=0)_{ch}
\label{chirel}
\eeq
Of course, the Ising model with zero field is not equivalent to one with
nonzero field, since in the former case the partition function and free energy
are exactly invariant under the $Z_2$ transformation $\sigma_n \to -\sigma_n$,
whereas in the latter case this symmetry is broken explicitly by the external
field term.  This inequivalence is manifested in the fact that eqs. 
(\ref{zrelation})-(\ref{k4p}) are not $Z_2$-invariant.  Thus under the
transformation $\sigma_n \to -\sigma_n$, which is equivalent to $h \to -h$, the
$i$'s in these equations are replaced by $-i$.  It is also manifested in the
fact that while the zero-field Ising model always has a $Z_2$-symmetric
paramagnetic (PM) phase, the model with nonzero external field $h \ne 0$ does
not have any PM phase.  The usefulness of eq. (\ref{frelation}) stems 
from the special feature that for $h=\pm i\pi/2$, this non-invariance is 
localised to just a constant term in the free energy. 

\vspace{2 mm}

The (reduced) free energy is \cite{ly} \footnote{The Hamiltonian in 
Ref. \cite{ly} was defined with a different zero point of the energy than that
used here.}
\beq
f(K,h= \pm i\pi/2) = \pm \frac{i \pi}{2} + \ln 2 + 
\frac{1}{4}\int_{-\pi}^{\pi}\int_{-\pi}^{\pi} \frac{d\theta_1 d\theta_2}
{(2\pi)^2} 
\ln \bigg \{ \frac{1}{2} \biggl [ C^4 + S^4 -1 + S^2 \Big ( \cos(\theta_1+
\theta_2) - \cos(\theta_1 - \theta_2) \Bigr ) \biggr ] \biggr \}
\label{fhgen}
\eeq
where $C$ and $S$ were defined in (\ref{cosh2k}) and (\ref{sinh2k}). 
From $U = -\partial f/\partial
\beta = -J \partial f/\partial K$, one has the symmetries 
\beq
U(\beta,J,H=\frac{i \pi}{2\beta}) = U(\beta,J,H=-\frac{i \pi}{2\beta})
\label{usymh}
\eeq
\beq
U(\beta,-J,h=\frac{i\pi}{2}) = U(\beta,J,h=\frac{i \pi}{2})
\label{usymj}
\eeq
\beq
U(-\beta,J,h=\frac{i\pi}{2}) = -U(\beta,J,h=\frac{i \pi}{2})
\label{usymbeta}
\eeq
Similarly, from $C=k_BK^2\partial^2f/\partial K^2$, one has
\beq
C(K,h=\frac{i\pi}{2}) = C(K,h=-\frac{i\pi}{2})
\label{csymh}
\eeq
and
\beq
C(K,h=\frac{i\pi}{2}) = C(-K,h=\frac{i\pi}{2})
\label{csymk}
\eeq
The free energy is trivially divergent at $K = \pm \infty$, i.e. $u=0$,
$\infty$; however, this will not be important here since these are isolated
points and not part of any phase boundaries.  The curves along which 
the free energy is non-analytic are given by the 
locus of points where the argument of the logarithm in the integrand of
eq. (\ref{fhgen}) vanishes.  Expressed in terms of the variable $u$, $f$ is 
\beq
f(K,h=\pm i\pi/2) = \pm \frac{i \pi}{2} + 
\frac{1}{4}\ln \biggl [ \frac{(1-u)^2}{u^2} \biggr ] 
+ \frac{1}{4}\int_{-\pi}^{\pi}\int_{-\pi}^{\pi} \frac{d\theta_1 d\theta_2}
{(2\pi)^2} \ln \biggl [(1+u)^2 - 2u P(\theta_1,\theta_2) \biggr ]
\label{fh}
\eeq
where 
\beq
P(\theta_1,\theta_2) = \cos \theta_1 + \cos \theta_2
\label{p}
\eeq
The above locus of points where the argument of the logarithm vanishes is given
by the solutions of the equation 
\beq
(1+u)^2 - 2ux = 0 
\label{ueq}
\eeq
where $x=P(\theta_1,\theta_2)$, taking values in the range $-2 \le x \le 2$. 
These are integrable singularities. 
Since the coefficients in this equation are real, the solutions are either real
or consist of complex conjugate pairs.  Moreover, under the replacement $u \to
1/u$, eq. (\ref{ueq}) retains its form, up to an overall factor of $u^{-2}$.
Consequently, the locus of solutions is also invariant under this mapping $u
\to 1/u$.  The solutions are shown in Fig. 2(a) and consist of the union of the
unit circle 
\beq
u = e^{i\phi} \ , \qquad -\pi < \phi \le \pi
\label{ucircle}
\eeq
and the finite line segment
\beq
\frac{1}{u_e} \le u \le u_e
\label{usegment}
\eeq
where the inner endpoint is 
\beq
u_e = -(3 - 2\sqrt{2}) = -0.171572875...
\label{ue}
\eeq
and the outer endpoint is $1/u_e = -(3+2\sqrt{2})=-5.828427...$. 
Note that $u_e=-u_c$, where $u_c$ is the usual critical point in the zero-field
square lattice Ising model separating the $Z_2$-symmetric, paramagnetic (PM) 
phase from the phase in in the $Z_2$ symmetry is spontaneously broken by
long-range ferromagnetic (FM) long-range order.

It is of interest to see how the solutions to eq. (\ref{ueq}) are traced out in
the complex $u$ plane as $x$ varies.
For $x=2$, this equation has a double root at $u=1$.  As $x$ decreases from
2 to 0, this root splits into a complex conjugate pair, the members of which
move counterclockwise and clockwise along the unit circle, and finally
rejoin to form a double root at $u=-1$ when $x=0$.  As $x$ decreases from 0
to $-2$, this double root again splits, but this time into two reciprocal real
roots, one of which moves to the right, from $u=-1$ to the endpoint $u_e$
and the other of which moves leftward to $u=1/u_e$. 
The corresponding phase boundaries in the $z$ plane consist of the unit
circle $|z|=1$ together with the two line segments from $z=\pm z_e=\pm
i(\sqrt{2}-1)$ upward and downward along the imaginary axis to $z = \mp 
1/z_e=\pm i(\sqrt{2}+1)$, respectively. 

The circle (\ref{ucircle}) divides the $u$ plane into two
separate phases.  A fundamental property of this model is that the nonzero
external field breaks the $Z_2$ symmetry explicitly, so that there is no
$Z_2$-symmetric phase.  Even without using the known expression 
for the magnetisation, one can identify the phases in this diagram as 
follows.  For sufficiently large real $K$, the interaction
of the external magnetic field with the spins is negligible compared with the 
spin-spin interaction, which thus produces a ferromagnetically ordered 
phase, just as it does in the model with $h=0$.  This shows that the
neighborhood of the origin in the $u$ (or $z$) plane is ferromagnetically
ordered.  By analytic continuation, it then follows that the entire region 
inside the unit circle $|u|=1$ is a ferromagnetically ordered phase, and
this is so denoted in Fig. 2(a).  Similarly, for sufficiently large negative 
$K$, the interaction of the external field with the spins is again negligible
compared with the spin-spin interaction, which produces a phase with 
antiferromagnetic (AFM) long-range order.  By analogous analytic continuation
arguments, it follows that the entire region outside the unit circle is the AFM
phase.  This may be shown as follows. 
One may think of the phase diagram in the complex variable $w=1/u$.
By the $u\to 1/u$ symmetry noted above, the phase boundaries in this variable
are the same as those in Fig. 2(a).  The above argument shows that the 
neighborhood of the origin is antiferromagnetically ordered, and hence, by the
same analytic continuation method, the entire region inside the unit circle
$|w|=1$ is an AFM phase.  Mapping this back to the $u$ plane, one has shown
that the entire region outside of the unit circle $|u|=1$ is the AFM
phase. Since the $u$ plane consists of precisely two regions with the property
that within each one can analytically continue from any one point to any other,
we have thus obtained a complete description of the phases in the model.  The 
same reasoning implies that in the $z$ plane, the phase with $|z| < 1$ is FM 
and the phase with $|z| > 1$ is AFM.   In passing, we note that 
complex-temperature properties of the $h=0$ Ising model on $d=2$ lattices 
have been studied in Refs. 
\cite{fisher,g69,dg,g75,ipz,ms,egj,chisq,chitri,chihc}.  In the case 
$h=0$ for the square lattice, the analogous locus of points across which the
free energy is singular form a lima\c{c}on \cite{chisq} defined by 
$Re(u) = 1 + 2^{3/2}\cos \omega + 2 \cos 2\omega$, 
$Im(u) = 2^{3/2}\sin \omega + 2 \sin 2\omega$ for $0 \le \omega < 2\pi$, or
equivalently, in the $z$ plane, the circles \cite{fisher} 
$z = \pm 1 + \sqrt{2} e^{i\theta}$, for $0 \le \theta < 2\pi$.

    By the use of the conformal mapping (\ref{bilinear}) on the $z$ plane, or
by re-expressing the free energy in terms of the variable $v$ and again solving
for the locus of points where the argument of the logarithm vanishes, we find
that the phase diagram of the model in the $v$ plane is as shown in Fig. 2(b). 
The unit circle $|z|=1$ is mapped to the imaginary axis in the $v$ plane, and 
the respective line segments from $z=\pm z_e$ to $\mp 1/z_e$ are mapped 
to the arcs from $v=e^{\mp i\pi/4}$ to $v=e^{\mp 3i\pi/4}$.  It is interesting
that the $Re(v)=0$ (i.e. imaginary) axis forms the boundary between the
complex-temperature FM and AFM phases.  One may understand this by recalling
that (i) if $h$ were real and positive (negative), this would favor FM (AFM) 
ordering, but a pure imaginary value of $h$ does not 
favor FM over AFM ordering, or vice versa; (ii) similarly, if the spin-spin 
coupling $K$ is real and positive (negative), it favors FM (AFM) ordering, but
a pure imaginary value of $K$ (and hence $v$) does not favor FM over AFM
ordering, or vice versa.  Therefore, if both $h$ and $K$ are pure imaginary, as
is the case here for the imaginary axis in the $v$ plane, then the system is
precisely balanced between FM and AFM order, so that this axis should be the
boundary between the complex-temperature FM and AFM phases, and this is just
what our explicit calculation shows.

    The mapping defined by $u \to \kappa^2$, where $\kappa$ was defined in
(\ref{kappa}),  takes the the locus of points 
(\ref{ucircle}) and (\ref{usegment}) to a single semi-infinite line segment 
extending from 1 to $\infty$ in the complex $\kappa^2$ plane. 
All points in the $\kappa^2$ plane are analytically connected to all other 
points.  In particular, the mapping $u \to \kappa^2$ takes both the 
complex-temperature FM and AFM phases in the $u$ plane to the same respective 
regions in the $\kappa^2$ plane, as is clear from the symmetry (\ref{uksym})
and the fact that the transformation $u \to 1/u$ interchanges the FM and AFM
phases in the $u$ plane.

\section{Complex-Temperature Behaviour of the Internal Energy and Specific 
Heat}
\label{uandc}

\subsection{Exact Expressions}
\label{expressions}

   From the free energy (\ref{fh}), it is straightforward to calculate the
internal energy $U$ and specific heat $C$ (per site).  In terms of the variable
$u$, we find that 
\beq
U = -J \biggl [ \ \frac{1+u}{1-u} + 
\Bigl ( \frac{1-u}{1+u}\Bigr ) \Bigl ( \frac{2}{\pi} \Bigr )K(\kappa) \ 
\biggr ]
\label{uh}
\eeq
where the elliptic modulus $\kappa$ was given above in eq. (\ref{kappa}) 
and $K(k)=\int_{0}^{\pi/2}(1-k^2\sin^2\theta)^{-1/2} d\theta$ is the 
complete elliptic integral of the first kind. This expression holds for both 
the FM and AFM phases and exhibits the symmetries (\ref{usymj}) and
(\ref{usymbeta}).  Since either of these has the effect of taking $u \to 1/u$,
and since this mapping takes the interior of the complex-temperature FM phase 
to the interior of the complex-temperature AFM phase, the values of $U$ in
these two phases are simply related by (\ref{usymj})-(\ref{usymbeta}). 
In the FM phase, the first few terms of the 
small-$|u|$ expansion (complex-temperature generalisation of the
low-temperature expansion) are 
\beq
U = -2J \Bigl [ 1+4u^2-12u^3+60u^4-280u^5 + O(u^6) \Bigr ]
\label{uexpansionfm}
\eeq
In the AFM phase, the corresponding expansion parameter is $w=1/u$, and $U$ has
the same expansion as (\ref{uexpansionfm}) with $J$ replaced by $-J$ and $u$
replaced by $w$.  

As discussed above, in the limit $J \to \infty$, and hence $K \to \infty$ with
fixed $H$, the spin-spin 
interaction overwhelms the contribution of the external field coupling, which
therefore has a negligible effect, to leading order.  It follows that in this 
limit, the value of the internal energy should be the same as the value for
$h=0$, i.e., 
\beq
U(u=0,h=i\pi/2) = U(u=0,h=0)
\label{uequality}
\eeq
It is interesting to compare the small-$|u|$ series expansions of these two
functions to ascertain the finite-$u$ corrections to this equality.  For this
purpose, we recall that \cite{ons}
\beq
U(K,h=0) = -J \biggl [ \ \frac{1+u}{1-u} + \frac{(1-6u+u^2)}{(1-u^2)}
\Bigl ( \frac{2}{\pi} \Bigr ) K(\kappa_0) \ \biggr ]
\label{u0}
\eeq
where 
\beq
\kappa_0 = \frac{4z(1-u)}{(1+u)^2}
\label{kappa0}
\eeq
The expression (\ref{c0}) holds for all phases, PM, FM, and AFM.  In the FM
phase, it has the small-$|u|$ expansion 
\beq
U(h=0) = -2J \Bigl [1-4u^2-12u^3-36u^4-120u^5 + O(u^6)  \Bigr ]
\label{u0expansionfm}
\eeq
Clearly, the expansions (\ref{uexpansionfm}) and (\ref{u0expansionfm}) agree
with the relation (\ref{uequality}) for $u=0$. 
$U(K,h=0)$ also satisfies the symmetries analogous to (\ref{usymj}) and 
(\ref{usymbeta}), with $h=i\pi/2$ replaced by $h=0$; as a consequence, in the
AFM phase, the small-$|w|$ expansion of $U(K,h=0)$ is given by
(\ref{u0expansionfm}) with $J$ replaced by $-J$ and $u$ replaced by $w$. 

\vspace{2mm}

  For $C$ we get
\beq
\frac{C}{8k_BK^2} = -\frac{u}{(1-u)^2} - \frac{(1+u)^2}{\pi(1+6u+u^2)}E(\kappa)
+ \frac{(1+u^2)}{\pi(1+u)^2}K(\kappa)
\label{ch}
\eeq
where $K(k)$ was defined above and
$E(k)=\int_{0}^{\pi/2}(1-k^2\sin^2\theta)^{1/2} d\theta$ is the complete
elliptic integral of the second kind. 
Again, this expression holds for both the FM and AFM phases.  It will also be
useful to express $C$ in an equivalent form, using (\ref{kappa}):
\beq
\frac{C}{8k_BK^2} = -\frac{u}{(1-u)^2} - \frac{E(\kappa)}{\pi(1+\kappa)} 
+ \frac{(1+u^2)(1+\kappa)K(\kappa)}{\pi(1+6u+u^2)}
\label{chkappa}
\eeq
$C/K^2$ has the small-$|u|$ expansion 
\beq
\frac{C}{8k_BK^2} = -64u^2 + 288u^3 - 1920u^4 + 11200u^5 + O(u^6)
\label{chexpansion}
\eeq
For comparison, the specific heat for the Ising model on the square lattice 
with $h=0$ \cite{ons} is
\beq
\frac{C}{k_BK^2} = \frac{4(1-\kappa')}{\pi \kappa^2}
\biggl [ 2 \{ K(\kappa_0)-E(\kappa_0) \} - (1-\kappa_0') \{ \frac{\pi}{2} + 
\kappa_0'K(\kappa_0) \} \biggr ]
\label{c0}
\eeq
which has the small-$|u|$ expansion
\beq
\frac{C}{k_BK^2} = 64u^2 + 288u^3 + 1152u^4 + 4800u^5 + O(u^6) 
\label{c0expansion}
\eeq
Of course, in this case, the positivity of the specific heat requires 
that the coefficient of the lowest order term must be
positive (the first negative coefficient occurs in the $u^7$ term). 
We proceed to determine the complex-temperature singularities of $U$ and $C$
for the present case, $h=i\pi/2$.  

\subsection{Vicinity of $u=u_e$}

   As discussed in connection with Fig. 2(a), the point $u=u_e$ is the
endpoint of the singular line segment protruding into the complex-temperature
extension of the FM phase.  All approaches to this point, except directly from
the left along this singular line segment, occur from within the
complex-temperature FM phase.  As $u \to u_e=-(3-2^{3/2})$, 
$\kappa \to -1$.  The internal energy diverges like 
\beq
\frac{U}{J} \to \frac{\sqrt{2}}{\pi}\ln(1-u/u_e) \ , \qquad as \ \ u \to u_e
\label{udivue}
\eeq
In the specific heat, the dominant divergence arises from the term in
(\ref{ch}) involving $E(\kappa)$ and is 
\beq
\frac{C}{k_BK^2} \to -\frac{4\sqrt{2}}{\pi(1-u/u_e)} \ , \qquad as \ \ u \to
u_e 
\label{cdivue}
\eeq
so that the associated singular exponent for $C$ at $u=u_e$ is 
\beq
\alpha_{e,FM}' \equiv \alpha_e' = 1
\label{alphae}
\eeq
where the subscript FM indicates the phase from which this point is approached,
and the prime is the standard notation indicating that the
approach to this singular point is from within a broken-symmetry phase.  (The
last feature is, of course, true of all of the singular points for $h \ne 0$.) 
There is also a weaker, logarithmic divergence arising from the term involving
$K(\kappa)$.  The value of $K$ at $u_e$ in (\ref{cdivue}) is
\beq
K_e=-\frac{1}{4}\ln u_e = -\frac{1}{4}\Bigl [ \ln(3-2^{3/2}) + i \pi + 2ni\pi
\Bigr ]
\label{ke}
\eeq
where $n$ labels the Riemann sheet used for the evaluation of the logarithm,
which we shall take to be $n=0$ below, unless otherwise indicated. 

\subsection{Vicinity of $u=1/u_e$}

  The point $u=1/u_e$ is the left end of the singular line segment protruding
into the complex-temperature extension of the AFM phase.  Except for the
approach directly from the right along the singular line segment, all
approaches to this point occur from within the complex-temperature AFM phase.
As $u \to 1/u_e$, $\kappa \to -1$, as is clear from the previous remarks and
the symmetry (\ref{uksym}).  The internal energy again diverges like 
\beq
\frac{U}{J} \to -\frac{\sqrt{2}}{\pi}\ln(1-u_eu) \ , \qquad as \ \ u \to
\frac{1}{u_e}
\label{udivuoe}
\eeq
In the specific heat, the dominant divergence again arises from the term in
(\ref{ch}) involving $E(\kappa)$ and is 
\beq
\frac{C}{k_BK^2} \to \frac{4\sqrt{2}}{\pi(1-u_eu)} \ , \qquad as \ \ u \to
1/u_e 
\label{cdivuoe}
\eeq
so that the associated singular exponent for $C$ at the outer endpoint (oe) 
$u=1/u_e$ is 
\beq
\alpha_{oe,AFM}' \equiv \alpha_{oe}' = 1
\label{alphaoe}
\eeq
The value of $K$ corresponding to $u=1/u_e$ in eq. (\ref{cdivuoe}) is, for the
principal Riemann sheet of the log, $K_{oe}=-(1/4)[\ln(3+2^{3/2})+i\pi]$. 

\subsection{Vicinity of $u=-1$}

   As $u \to -1$ (denoted $u_s$), $\kappa$ diverges; if we set $u=-1+\epsilon
e^{i \phi}$ and let $\epsilon \to 0$, then 
$\kappa \sim -4 \epsilon^{-2}e^{-2 i \phi}$.  One easily sees that $U(u=-1)$ is
finite.  For $C$, we observe that the first and second terms on the right-hand
side of eq. (\ref{ch}) or (\ref{chkappa}) are finite.  By the use of the 
elliptic integral identity $(1+\kappa)K(\kappa)=K(2\kappa^{1/2}/(1+\kappa))$ 
(see, e.g. \cite{grad}), we can rewrite the term involving $K(\kappa)$ in
eq. (\ref{chkappa}) as
\beq
\frac{(1+u^2)}{\pi(1+6u+u^2)}K\Bigl ( \frac{2\kappa^{1/2}}{1+\kappa} \Bigr )
\label{kterm}
\eeq
As $u\to -1$ and $\kappa$ diverges, $K(2\kappa^{1/2}/(1+\kappa)) \to
K(0)=\pi/2$, so that $C$ is finite, although non-analytic, at $u=-1$.  This is
true for the approach to $u=-1$ from either the FM or AFM phases. We thus 
have 
\beq
\alpha_{s,FM}'=\alpha_{s,AFM}'= 0 \qquad (log. \ finite)
\label{alphas}
\eeq

\subsection{Vicinity of $u=1$}

   As $u \to 1$, $\kappa \to 1$. 
In $U$ the leading potential singularity arises from the first term in 
(\ref{uh})
\beq 
\frac{U}{J} \to \frac{-2}{1-u} \qquad as \ \ u \to 1
\label{udiv0}
\eeq 
Now $K=-(1/4)\ln u$, so that, if one uses the first Riemann sheet of the
logarithm, then $u \to 1$ maps to $K \to 0$.  Recalling that $K=\beta J$, if
this zero in $K$ is due to $\beta \to 0$ at fixed nonzero $J$, then
eq. (\ref{udiv0}) shows that $U$ diverges for $u \to 1$; however, if the 
zero in $K$ is due to $J \to 0$ at fixed nonzero $\beta$, then, expanding 
(\ref{udiv0}), one finds that $U \to 1/(2\beta)$.  

   For the specific heat, from (\ref{ch}), it follows that as $u \to 1$, 
\beq
k_B^{-1}C \to 8K^2 \biggl [ -\frac{1}{(1-u)^2} + \frac{1}{4\pi}\ln 
\Bigl ( \frac{32}{(1-u)^2} \Bigr ) + ... \biggr ]
\label{cnear1}
\eeq
where $...$ refers to less singular terms.  If we again use the principal
Riemann sheet of the logarithm, so that $u \to 1$ corresponds to $K \to 0$, 
then (\ref{cnear1}) becomes 
\beq
k_B^{-1}C \to -\frac{1}{2} + \frac{2}{\pi}K^2 \ln \Bigl (\frac{2}{K^2} \Bigr )
+ O(K^2) \qquad \rm{as} \ u \to 1
\label{cnear1k}
\eeq
That is, $C$ has a finite logarithmic singularity at this point, and hence a
corresponding exponent
\beq
\alpha_{1,FM}' = \alpha_{1,AFM}'=0  \qquad (log. \ finite)
\label{alpha1}
\eeq
In passing, we note that if one were to use a Riemann sheet different than the
principal ($n=0$) one in evaluating $K = -(1/4)\ln(1)$, so that $K=-i n \pi/2
\ne 0$, then $C$ would diverge quadratically at $u=1$.  

\subsection{Elsewhere Along the Singular Curves}

   We discuss here the behaviour of $U$ and $C$ as one crosses the singular 
locus of points comprised by the unit circle (\ref{ucircle}) and the line
segment (\ref{usegment}) away from the points $u=u_e$, $1/u_e$, $-1$, and 1.
The singularities which one encounters in this case are associated with passage
across the branch cut of the elliptic integrals in (\ref{uh}) and
(\ref{ch}).  We recall that the elliptic integrals $K(\kappa)$ and $E(\kappa)$
are analytic functions of $\kappa^2$ in the complex $\kappa^2$ plane except for
respectively divergent and finite branch points at $\kappa^2=1$ and an
associated branch cut, which is normally taken to run from $\kappa^2=1$ to
$\kappa^2=\infty$ along the positive real axis in this plane.  To illustrate
the nature of the singularities, we shall consider moving outward along a 
ray in the $u$ plane defined by $u=\rho e^{i\theta}$ with $\rho$ increasing 
from 0 to $\infty$ at fixed $\theta$, say $\theta=\pi/6$.  
As shown in Fig. 3, the image point in the $\kappa^2$ plane also moves out 
from the origin, starting with an angle of $\pi/3$ but bending around to the
right.  As we cross the unit circle in the $u$ plane, leaving the FM phase and
entering the AFM phase, the image point in the $\kappa^2$ plane crosses the
branch cut moving vertically downward.  This branch cut is precisely the image
of the unit circle $|u|=1$ (and also of the singular line segment
(\ref{usegment}).) For $\theta=\pi/6$, the crossing point is
at $\kappa^2=2^4(7-4\sqrt{3})=1.1487...$.  In the $\kappa^2$ plane, one thus
passes onto the second Riemann sheet of the elliptic functions $K(\kappa)$ and
$E(\kappa)$.  If one projects back to the first Riemann sheet, these functions
have discontinuous imaginary parts across this branch cut.  As $\rho$ continues
to increase toward $\infty$, $\kappa^2 \sim 16\rho^{-2}e^{-2\theta}$ so that 
the image point curves around finally approaches the origin in a ``northwest''
direction, at an angle of $-\pi/3$, but on the second 
Riemann sheet.  In Fig. 3 we show the image point for $\rho$ in the range 
from 0 to 30.

\section{Complex-Temperature Behaviour of the Uniform and Staggered 
Magnetisation }

   The magnetisation $M$ is \cite{ly,mw}
\beq
M(u,h=i\pi/2) = \frac{(1+u)^{1/2}}{(1-u)^{1/4}(1+6u+u^2)^{1/8}}
\label{mh}
\eeq
Note that $(1+6u+u^2) = (1-u/u_e)(1-u_eu)$.  By
analytic continuation, this formula holds throughout the complex-temperature
extension of the FM phase.  The identity discussed above, and the resultant 
eq. (\ref{magrel}) yields the relation
\beq
M(u,h=\pm i \pi/2) = M(-u,h=0)^{-1}
\label{mrelexplicit}
\eeq
where \cite{yang}
\beq
M(u,h=0) = \frac{(1+u)^{1/4}(1-6u+u^2)^{1/8}}{(1-u)^{1/2}}
\label{msq}
\eeq
As is well known, one can express $M(u,h=0)$ as 
\beq
M(u,h=0) = (1-(k_{<,0})^2)^{1/8}
\label{mk}
\eeq
where 
\beq
k_{<,0} = \frac{1}{\sinh^2(2K)} = \frac{4u}{(1-u)^2}
\label{klsq}
\eeq 
This quantity also enters in exact expressions for correlation functions in the
FM phase of the $h=0$ Ising model \cite{ok,mpw,gs}. 
Given the relation (\ref{mrelexplicit}), it is natural to write 
\beq
M(u,h=i\pi/2) = (1-(k_<)^2)^{1/8}
\label{mkh}
\eeq
where the elliptic modulus was introduced in eq. (\ref{kl}). 
The magnetisation for the $h=i\pi/2$ case vanishes continuously at the point
$u=-1$ (denoted $u_s$) with exponent 
\beq
\beta_s = \frac{1}{2}
\label{betas}
\eeq
diverges at $u=u_e$ with exponent
\beq
\beta_e = -\frac{1}{8}
\label{betae}
\eeq
and diverges at $u=1$ with exponent
\beq
\beta_1 = -\frac{1}{4}
\label{beta1}
\eeq
Elsewhere on the boundary of the complex-temperature extension of the FM phase,
i.e., the unit circle in the $u$ plane, $M$ vanishes discontinously.  Note that
the apparent divergence at the point $u=1/u_e$ does not actually
occur, since this is outside of the complex-temperature FM phase, where the
above analytic continuation is valid.  

   The staggered magnetisation $M_{st}$ does not seem to have been explicitly 
discussed in the literature, but one can easily obtain it, as follows.
$M_{st}$ may be defined via 
\beq
M_{st}^2 = \lim_{|{\bf r}| \to \infty} 
<\tilde \sigma_{ \bf 0} \tilde \sigma_{\bf r}>
\label{mstsq}
\eeq
where
\beq
\tilde \sigma_{\bf r} = (-1)^{p({\bf r})}\sigma_{\bf r}
\label{sigmatilde}
\eeq
where
\beq
p({\bf r}) = {\sum_{i=1}^{2} r_i}
\label{pfun}
\eeq
i.e. $\tilde \sigma_{\bf r} = \sigma_{\bf r}$ for $\bf r$ on the same 
sublattice as $\bf r = \bf 0$ and $-\sigma_{\bf r}$ for $\bf r$ on the
other sublattice of the (bipartite) square lattice.  To evaluate $M_{st}$ via
eq. (\ref{mstsq}), it suffices to take ${\bf r} = (r,0)$ or $(0,r)$, i.e. the
row or column 2-spin correlation function.  From the known asymptotic 
behaviour of this correlation function \cite{mw}, one immediately finds that 
\beq
M_{st}(w) = M(u \to w)
\label{mmst}
\eeq
where, as before, $w=1/u$. This is consistent with
eq. (\ref{mkh}) since (c.f. (\ref{uksym})) $u \to 1/u$ takes
$k_< \to -k_<$, and $k_<$ enters squared in (\ref{mkh}). Of course, $M_{st}$ 
vanishes identically outside the complex-temperature extension of the AFM
phase.  Further, we may immediately conclude that $M_{st}$ vanishes
continuously at $u=-1$ with exponent (\ref{betas}), diverges at $u=1/u_e$ with
exponent (\ref{betae}), and diverges at $u=1$ with exponent (\ref{beta1}). 
Elsewhere along the boundary of the complex-temperature AFM phase, $M_{st}$ 
vanishes discontinuously, with the same discontinuity as $M$. 

    It is of interest to compare these results with the behaviour of $M$ and 
$M_{st}$ for $h=0$ (again on the square lattice).  Aside from the physical 
PM-FM and PM-AFM critical points $u=u_c=(3-2^{3/2})$ and $1/u_c$, where,
respectively, $M$ and $M_{st}$ vanish continuously with exponent $\beta=1/8$,
they also both vanish continuously at the complex-temperature point $u=-1$,
with the same exponent, $\beta=1/4$.  Note that for $h=0$ there is only one 
point, viz., $u=-1$, where the FM and AFM phases are contiguous and $M$ and 
$M_{st}$ vanish continuously, whereas for $h=i\pi/2$ there are two such 
points, namely, $u=-1$ and $u=1$. 

\section{Extraction and Analysis of the Low-Temperature Series for $\bar\chi$}
\label{analysis}

\subsection{Generalities}
\label{general_low}

   In order to investigate the complex-temperature singularities of the
susceptibility $\bar\chi$, we shall make use of the low-temperature, 
high-field series expansion for the free energy or equivalently the partition 
function of the Ising model on the square lattice \cite{tlow1,tlow2,be}. In
Ref. \cite{be}, Baxter and Enting calculated this expansion for the partition 
function to order $O(u^{23})$.  The series for $Z_r$ in eq. (\ref{zr}) is 
\beq
Z_r = 1 + \sum_{n=2}^{\infty}\sum_{m} a_{n,m}u^n \mu^m
\label{zrsq}
\eeq
where $j \le m \le j^2$ for $n=2j$ and $j \le m \le j(j-1)$ for $n=2j-1$.  We
extract the series for $h=i\pi/2$ by calculating $\bar\chi=\partial^2
f/\partial h^2$ and then subsituting $\mu=-1$. This has the form 
\beq
\bar\chi(u,h=i\pi/2) = 4u^2(\sum_{n=0}^{\infty} c_n u^n )
\label{chihuseries}
\eeq
The results for the $c_n$ are listed in Table 1; the series for $Z$ and the
resultant series for 
$\bar\chi$ to $O(u^{23})$ yields the $c_n$'s to order $n=21$.  
Parenthetically, we note that $\bar\chi$ has been 
calculated to $O(u^{28})$ in Ref. \cite{beg} and to $O(u^{38})$ in
Ref. \cite{egj}, but the low-temperature, high-field expansion of
the partition function as a function of $\mu=e^{-2h}$ which would be 
necessary to extract $\bar\chi(h=i\pi/2)$ was not given in 
these papers (it would be a rather long expression). 

\begin{table}
\begin{center}
\begin{tabular}{|c|c|} \hline \hline
$n$ & $c_n$ \\ 
\hline \hline
0   &   $-1$                \\ \hline
1   &   8                   \\ \hline
2   &   $-48$               \\ \hline
3   &   304                 \\ \hline
4   &   $-1863$             \\ \hline
5   &   11368               \\ \hline
6   &   $-68840$            \\ \hline
7   &   414872              \\ \hline
8   &   $-2490437$          \\ \hline
9   &   14903648            \\ \hline
10  &   $-88963696$         \\ \hline
11  &   529939176           \\ \hline
12  &   $-3151205475$       \\ \hline
13  &   18710180192         \\ \hline
14  &   $-110948037424$     \\ \hline
15  &   657164715520        \\ \hline
16  &   $-3888670886593$    \\ \hline
17  &   22990566432904      \\ \hline
18  &   $-135819110416784$  \\ \hline
19  &   801806651588848     \\ \hline 
20  &   $-4730485389238263$ \\ \hline
21  &   27892958533539784   \\ \hline
\hline
\end{tabular}
\end{center}
\caption{Low-temperature series expansion coefficients for
$\bar\chi(u,h=i\pi/2)$ in eq. (\ref{chihuseries}).}
\label{table1}
\end{table}
We have analysed this series using dlog Pad\'{e} and differential 
approximants.  For a recent review of these techniques, see Ref. 
\cite{tonyg}. Our notation for these approximants follows Ref. \cite{tonyg} and
our earlier work on complex-temperature properties of the $h=0$ Ising model 
\cite{chisq,chitri,chihc}.  In particular, we use first order differential 
approximants (i.e., $K=1$ in our previous notation); as before, we used
unbiased approximants so as to be able to use an extrapolation method for
extracting critical exponents.  
Since the prefactor $4u^2$ is analytic, we have actually
performed the analysis on the reduced function
\beq
\bar\chi_r \equiv \frac{\bar\chi}{4u^2} = 
\sum_{n=0}^{\infty} c_n u^n
\label{chir}
\eeq
As one approaches a generic complex singular point denoted 
$sing$ from within the complex-temperature extension of the FM phase,  
$\bar\chi$ is assumed to have the leading singularity 
\beq
\bar\chi \sim A_{sing}'(1-u/u_{sing})^{-\gamma_{sing}'}
\Bigl ( 1+a_1(1-u/u_{sing}) + ... \Bigr )
\label{singform}
\eeq
where $A_{sing}'$ and $\gamma_{sing}'$ denote, respectively, the critical 
amplitude and
the corresponding critical exponent, and the $...$ represent analytic confluent
corrections.  One may observe that we have not included non-analytic confluent
corrections to the scaling form in eq. (\ref{singform}).  The reason is that,
as discussed in our earlier work \cite{chisq}, previous studies have 
indicated that they are very weak or absent for the 2D Ising model. 
We proceed to our results. 

\subsection{Singularity at $u=u_e$}

    Our dlog Pad\'{e} results relevant to the singularity in 
$\bar\chi(u,h=i\pi/2)$ at $u=u_e$ are given in Table 2. 
\begin{table}
\begin{center}
\begin{tabular}{|c|c|c|c|} \hline \hline & & & \\ 
$[N/D]$ & $u_{sing}$ & $|u_{sing}-u_e|/|u_e|$ & $\gamma_e'$ \\
& & & \\
\hline \hline
$[5/3]$ & $-0.1716072$ & $2.0 \times 10^{-4} $ & $    1.2492$ \\ \hline
$[4/4]$ & $-0.1715705$ & $1.4 \times 10^{-5} $ & $    1.2464$ \\ \hline
$[5/4]$ & $-0.1715636$ & $5.4 \times 10^{-5} $ & $    1.2459$ \\ \hline
$[6/4]$ & $-0.1715646$ & $4.8 \times 10^{-5} $ & $    1.2460$ \\ \hline
$[3/5]$ & $-0.1715802$ & $4.3 \times 10^{-5} $ & $    1.2471$ \\ \hline
$[4/5]$ & $-0.1715628$ & $5.8 \times 10^{-5} $ & $    1.2458$ \\ \hline
$[5/5]$ & $-0.1715645$ & $4.9 \times 10^{-5} $ & $    1.2460$ \\ \hline
$[6/5]$ & $-0.1715598$ & $7.6 \times 10^{-5} $ & $    1.2457$ \\ \hline
$[7/5]$ & $-0.1715670$ & $3.4 \times 10^{-5} $ & $    1.2462$ \\ \hline
$[4/6]$ & $-0.1715647$ & $4.8 \times 10^{-5} $ & $    1.2460$ \\ \hline
$[5/6]$ & $-0.1715674$ & $3.2 \times 10^{-5} $ & $    1.2462$ \\ \hline
$[6/6]$ & $-0.1715667$ & $3.6 \times 10^{-5} $ & $    1.2462$ \\ \hline
$[7/6]$ & $-0.1715667$ & $3.6 \times 10^{-5} $ & $    1.2462$ \\ \hline
$[8/6]$ & $-0.1715665$ & $3.7 \times 10^{-5} $ & $    1.2462$ \\ \hline
$[5/7]$ & $-0.1715668$ & $3.6 \times 10^{-5} $ & $    1.2462$ \\ \hline
$[6/7]$ & $-0.1715667$ & $3.6 \times 10^{-5} $ & $    1.2462$ \\ \hline
$[7/7]$ & $-0.1715670$ & $3.4 \times 10^{-5} $ & $    1.2462$ \\ \hline
$[8/7]$ & $-0.1715664$ & $3.8 \times 10^{-5} $ & $    1.2461$ \\ \hline
$[9/7]$ & $-0.1715664$ & $3.7 \times 10^{-5} $ & $    1.2461$ \\ \hline
$[6/8]$ & $-0.1715672$ & $3.3 \times 10^{-5} $ & $    1.2462$ \\ \hline
$[7/8]$ & $-0.1715664$ & $3.8 \times 10^{-5} $ & $    1.2461$ \\ \hline
$[8/8]$ & $-0.1715665$ & $3.7 \times 10^{-5} $ & $    1.2461$ \\ \hline
$[9/8]$ & $-0.1715754$ & $1.5 \times 10^{-5} $ & $    1.2501$ \\ \hline
$[10/8]$ & $-0.1715673$ & $3.2 \times 10^{-5} $ & $    1.2463$ \\ \hline
$[7/9]$ & $-0.1715665$ & $3.7 \times 10^{-5} $ & $    1.2461$ \\ \hline
$[8/9]$ & $-0.1715666$ & $3.6 \times 10^{-5} $ & $    1.2462$ \\ \hline
$[9/9]$ & $-0.1715673$ & $3.2 \times 10^{-5} $ & $    1.2463$ \\ \hline
$[10/9]$ & $-0.1715677$ & $3.0 \times 10^{-5} $ & $    1.2464$ \\ \hline
$[11/9]$ & $-0.1715696$ & $1.9 \times 10^{-5} $ & $    1.2470$ \\ \hline
$[8/10]$ & $-0.1715674$ & $3.2 \times 10^{-5} $ & $    1.2463$ \\ \hline
$[9/10]$ & $-0.1715686$ & $2.5 \times 10^{-5} $ & $    1.2466$ \\ \hline
$[10/10]$ & $-0.1715699$ & $1.7 \times 10^{-5} $ & $    1.2471$ \\ \hline
$[9/11]$ & $-0.1715701$ & $1.6 \times 10^{-5} $ & $    1.2472$ \\ \hline 
\hline
\end{tabular}
\end{center}
\caption{Values of pole near $u_e=-(3-2^{3/2})= -0.171572875...$, normalised
distance from this point, $|u_{sing}-u_e|/|u_e|$, and exponent $\gamma_e'$ from
dlog Pad\'{e} approximants to low-temperature series for 
$\bar\chi_r(u,h=i\pi/2)$, starting
with the series to $O(u^9)$.  We list 
only the approximants which satisfy the accuracy criterion 
$|u_{sing}-u_e|/|u_e| \le 2 \times 10^{-4}$.}
\label{table2}
\end{table}
Because of the length of our differential approximant results relevant
to the singularity in $\bar\chi$ at $u=u_e$, we list these in the Appendix. 
From these results, we infer the location of the singularity to be 
\beq
u_{sing.} = -0.17157 \pm 0.00001
\label{ueser}
\eeq
This, together with our knowledge of the exact location of the endpoint of 
the singular line segment (\ref{usegment}), supports the conclusion that the
exact location of this singularity is at
$u=u_e=-0.1715729...$ given in (\ref{ue}). 
Accepting this conclusion, we plot the values 
of the corresponding exponent $\gamma_e'$ for the differential approximants 
as functions of the normalised distance from this point, i.e., 
$|u_{sing.}-u_e|/|u_e|$ and extrapolate to zero distance.  (This extrapolation
method is similar to the use of biased differential approximants; in both of
these approaches, one uses one's knowledge of the exact position of the
singularity.)  From our extrapolation, we obtain
\beq
\gamma_e' = 1.25 \pm 0.01
\label{gammaeser}
\eeq
This strongly supports the following inference for the exact value of this
exponent, which we shall make: 
\beq
\gamma_e' = \frac{5}{4}
\label{gammae}
\eeq

   To calculate the critical amplitude for $\bar\chi$ at the $u=u_e$ 
singularity, we use the standard method of analysing Pad\'{e} approximants 
to the series $(-\bar\chi_r)^{1/\gamma_e'}$ (where the minus sign is inserted 
because $c_0=-1$ in eq. (\ref{chihuseries})).  For $A_e'$ as defined in
eq. (\ref{singform}) with $u_{sing}=u_e$, we obtain 
\beq
A_e'=  -0.11515 \pm 0.00020
\label{ampe}
\eeq

\subsection{Singularity at $u=1$}

   Our dlog Pad\'{e} results for $\bar\chi_r$ relevant to the singularity at
$u=1$ are given in Table 3.
We obtain more precise results from differential approximants; these are
given in Table 4 and its continuation. 
\begin{table}
\begin{center}
\begin{tabular}{|c|c|c|c|} \hline \hline  & & & \\
$[N/D]$ & $u_{sing}$ & $|u_{sing}-1|$ & $\gamma_1'$ \\
 & & & \\
\hline \hline
$[4/5]$ & $ 1.0024444$ & $2.4 \times 10^{-3} $ & $    2.3646$ \\ \hline
$[6/6]$ & $ 0.9894269$ & $1.1 \times 10^{-2} $ & $    2.3347$ \\ \hline
$[5/7]$ & $ 0.9920029$ & $8.0 \times 10^{-3} $ & $    2.3623$ \\ \hline
$[6/7]$ & $ 0.9817550$ & $1.8 \times 10^{-2} $ & $    2.2642$ \\ \hline
$[7/7]$ & $ 0.9878026$ & $1.2 \times 10^{-2} $ & $    2.3213$ \\ \hline
$[8/7]$ & $ 0.9988081$ & $1.2 \times 10^{-3} $ & $    2.4483$ \\ \hline
$[6/8]$ & $ 0.9892965$ & $1.1 \times 10^{-2} $ & $    2.3380$ \\ \hline
$[8/8]$ & $ 0.9922151$ & $7.8 \times 10^{-3} $ & $    2.3641$ \\ \hline
$[9/8]$ & $ 0.9969711$ & $3.0 \times 10^{-3} $ & $    2.4255$ \\ \hline
$[10/8]$ & $ 0.9974836$ & $2.5 \times 10^{-3} $ & $    2.4332$ \\ \hline
$[7/9]$ & $ 0.9934910$ & $6.5 \times 10^{-3} $ & $    2.3814$ \\ \hline
$[8/9]$ & $ 1.0054945$ & $5.5 \times 10^{-3} $ & $    2.5876$ \\ \hline
$[9/9]$ & $ 0.9975390$ & $2.5 \times 10^{-3} $ & $    2.4342$ \\ \hline
$[10/9]$ & $ 0.9967646$ & $3.2 \times 10^{-3} $ & $    2.4231$ \\ \hline
$[11/9]$ & $ 0.9982017$ & $1.8 \times 10^{-3} $ & $    2.4446$ \\ \hline
$[8/10]$ & $ 0.9980098$ & $2.0 \times 10^{-3} $ & $    2.4427$ \\ \hline
$[9/10]$ & $ 0.9992196$ & $7.8 \times 10^{-4} $ & $    2.4648$ \\ \hline
$[10/10]$ & $ 0.9984417$ & $1.6 \times 10^{-3} $ & $    2.4495$ \\ \hline
$[9/11]$ & $ 0.9985604$ & $1.4 \times 10^{-3} $ & $    2.4519$ \\ \hline
\hline \hline
\end{tabular}
\end{center}
\caption{Values of pole near $u=1$, normalised distance from
this point, $|u_{sing}-1|$, and exponent $\gamma_1'$ from $[M/N]$ dlog Pad\'{e}
approximants to low-temperature series for $\bar\chi_r$.  We list
only the differential approximants which satisfy the accuracy criterion
$|u_{sing}-1| < 2 \times 10^{-2}$.}
\label{table3}
\end{table}
\begin{table}
\begin{center}
\begin{tabular}{|c|c|c|c|} \hline \hline  & & & \\
$[L/M_0;M_1]$ & $u_{sing}$ & $|u_{sing}-1|$ & $\gamma_1'$ \\
 & & & \\
\hline \hline
$[0/7;7]$ & $ 1.0019932$ & $2.0 \times 10^{-3} $ & $   2.5262$ \\ \hline
$[0/7;9]$ & $ 0.9983638$ & $1.6 \times 10^{-3} $ & $   2.4478$ \\ \hline
$[0/8;9]$ & $ 0.9974253$ & $2.6 \times 10^{-3} $ & $   2.4337$ \\ \hline
$[0/8;10]$ & $ 0.9981341$ & $1.9 \times 10^{-3} $ & $  2.4447$ \\ \hline
$[0/9;9]$ & $ 0.9999105$ & $0.89 \times 10^{-4} $ & $  2.4776$ \\ \hline
$[0/9;10]$ & $ 0.9984608$ & $1.5 \times 10^{-3} $ & $  2.4502$ \\ \hline
$[0/10;9]$ & $ 0.9982548$ & $1.7 \times 10^{-3} $ & $  2.4461$ \\ \hline
$[1/4;5]$ & $ 1.0013165$ & $1.3 \times 10^{-3} $ & $   2.4393$ \\ \hline
$[1/6;8]$ & $ 0.9989524$ & $1.0 \times 10^{-3} $ & $   2.4554$ \\ \hline
$[1/7;9]$ & $ 0.9981225$ & $1.9 \times 10^{-3} $ & $   2.4445$ \\ \hline
$[1/8;8]$ & $ 1.0019138$ & $1.9 \times 10^{-3} $ & $   2.5153$ \\ \hline
$[1/8;9]$ & $ 0.9982128$ & $1.8 \times 10^{-3} $ & $   2.4460$ \\ \hline
$[1/8;10]$ & $ 0.9981111$ & $1.9 \times 10^{-3} $ & $  2.4444$ \\ \hline
$[1/9;9]$ & $ 0.9990590$ & $0.94 \times 10^{-3} $ & $  2.4617$ \\ \hline
$[2/4;6]$ & $ 1.0004562$ & $0.46 \times 10^{-3} $ & $  2.4545$ \\ \hline
$[2/6;4]$ & $ 1.0008071$ & $0.81 \times 10^{-3} $ & $   2.4534$ \\ \hline
$[2/6;6]$ & $ 1.0012170$ & $1.2 \times 10^{-3} $ & $   2.4958$ \\ \hline
$[2/6;7]$ & $ 0.9997299$ & $2.7 \times 10^{-4} $ & $   2.4678$ \\ \hline
$[2/6;8]$ & $ 0.9973495$ & $2.7 \times 10^{-3} $ & $   2.4543$ \\ \hline
$[2/7;6]$ & $ 0.9996006$ & $4.0 \times 10^{-4} $ & $   2.4651$ \\ \hline
$[2/7;9]$ & $ 0.9981442$ & $1.9 \times 10^{-3} $ & $   2.4450$ \\ \hline
$[2/8;8]$ & $ 1.0006495$ & $0.65 \times 10^{-3} $ & $  2.4969$ \\ \hline
$[2/8;9]$ & $ 0.9985512$ & $1.4 \times 10^{-3} $ & $   2.4513$ \\ \hline
$[2/9;8]$ & $ 0.9983746$ & $1.6 \times 10^{-3} $ & $   2.4474$ \\ \hline
$[3/6;6]$ & $ 1.0005153$ & $0.52 \times 10^{-3} $ & $  2.4841$ \\ \hline
$[3/6;7]$ & $ 0.9973962$ & $2.6 \times 10^{-3} $ & $   2.4232$ \\ \hline
$[3/6;8]$ & $ 0.9974009$ & $2.6 \times 10^{-3} $ & $   2.4201$ \\ \hline
$[3/7;6]$ & $ 0.9971786$ & $2.8 \times 10^{-3} $ & $   2.4189$ \\ \hline
$[3/7;7]$ & $ 0.9977559$ & $2.2 \times 10^{-3} $ & $   2.4302$ \\ \hline
$[3/7;8]$ & $ 0.9981732$ & $1.8 \times 10^{-3} $ & $   2.4407$ \\ \hline
$[3/7;9]$ & $ 0.9980242$ & $2.0 \times 10^{-3} $ & $   2.4651$ \\ \hline
$[3/8;6]$ & $ 0.9975202$ & $2.5 \times 10^{-3} $ & $   2.4227$ \\ \hline
\hline
\end{tabular}
\end{center}
\caption{Values of pole near $u=1$, normalised distance from
this point, $|u_{sing}-1|$, and exponent $\gamma_1'$ from
differential approximants to low-temperature series for $\bar\chi_r$.  We list
only the differential approximants which satisfy the accuracy criterion
$|u_{sing}-1| \le 2 \times 10^{-3}$.}
\label{table4}
\end{table}
\begin{table}
\begin{center}
\begin{tabular}{|c|c|c|c|} \hline \hline 
$[3/8;7]$ & $ 0.9980148$ & $2.0 \times 10^{-3} $ & $   2.4375$ \\ \hline
$[3/8;8]$ & $ 0.9986544$ & $1.3 \times 10^{-3} $ & $   2.4508$ \\ \hline
$[3/9;7]$ & $ 0.9991902$ & $0.81 \times 10^{-3} $ & $  2.4621$ \\ \hline
$[4/4;4]$ & $ 1.0024665$ & $2.5 \times 10^{-3} $ & $   2.4681$ \\ \hline
$[4/5;7]$ & $ 0.9979715$ & $2.0 \times 10^{-3} $ & $   2.4379$ \\ \hline
$[4/6;8]$ & $ 0.9998460$ & $1.5 \times 10^{-4} $ & $   2.4832$ \\ \hline
$[4/7;5]$ & $ 0.9982928$ & $1.7 \times 10^{-3} $ & $   2.4505$ \\ \hline
$[4/7;7]$ & $ 0.9989074$ & $1.1 \times 10^{-3} $ & $   2.4584$ \\ \hline
$[4/7;8]$ & $ 0.9990659$ & $0.93 \times 10^{-3} $ & $  2.4624$ \\ \hline
$[4/8;7]$ & $ 0.9990450$ & $0.96 \times 10^{-3} $ & $  2.4618$ \\ \hline
$[5/4;5]$ & $ 1.0009762$ & $0.98 \times 10^{-3} $ & $  2.4429$ \\ \hline
$[5/5;7]$ & $ 0.9987068$ & $1.3 \times 10^{-3} $ & $   2.4543$ \\ \hline
$[5/6;7]$ & $ 0.9993881$ & $0.61 \times 10^{-3} $ & $  2.4709$ \\ \hline
$[5/6;8]$ & $ 0.9991317$ & $0.87 \times 10^{-3} $ & $  2.4644$ \\ \hline
$[5/7;6]$ & $ 1.0014642$ & $1.5 \times 10^{-3} $ & $   2.5280$ \\ \hline
$[5/7;7]$ & $ 0.9990811$ & $0.92 \times 10^{-3} $ & $  2.4629$ \\ \hline
$[5/8;6]$ & $ 1.0000180$ & $1.8 \times 10^{-5} $ & $   2.4888$ \\ \hline
$[6/4;6]$ & $ 0.9993653$ & $0.63 \times 10^{-3} $ & $  2.5040$ \\ \hline
$[6/5;7]$ & $ 1.0002641$ & $2.6 \times 10^{-4} $ & $   2.4944$ \\ \hline
$[6/6;7]$ & $ 0.9991189$ & $0.88 \times 10^{-3} $ & $  2.4641$ \\ \hline
$[6/7;5]$ & $ 1.0003178$ & $3.2 \times 10^{-4} $ & $   2.4956$ \\ \hline
$[6/7;6]$ & $ 0.9993080$ & $0.69 \times 10^{-3} $ & $  2.4698$ \\ \hline
$[7/5;6]$ & $ 0.9990014$ & $1.0 \times 10^{-3} $ & $   2.4565$ \\ \hline
$[7/6;5]$ & $ 0.9983233$ & $1.7 \times 10^{-3} $ & $   2.4405$ \\ \hline
$[8/4;6]$ & $ 0.9993253$ & $0.67 \times 10^{-3} $ & $  2.4757$ \\ \hline
$[8/5;5]$ & $ 0.9987939$ & $1.2 \times 10^{-3} $ & $   2.4448$ \\ \hline
$[8/5;6]$ & $ 0.9994784$ & $0.52 \times 10^{-3} $ & $  2.4852$ \\ \hline
$[8/6;5]$ & $ 0.9981646$ & $1.8 \times 10^{-3} $ & $   2.4390$ \\ \hline
$[9/4;6]$ & $ 0.9986630$ & $1.3 \times 10^{-3} $ & $   2.4600$ \\ \hline
\hline \hline
\end{tabular}
\end{center}
\caption{Continuation of table of differential approximants for the singularity
in $\bar\chi_r$ at $u=1$.}
\label{table4a}
\end{table}
From these results, we obtain the location of the singularity as 
\beq
u_{sing} = 0.999 \pm 0.001
\label{u1ser}
\eeq
From this and our determination of the phase boundaries
(\ref{ucircle})-(\ref{usegment}), we infer that the exact location of this
singularity is at $u=1$.  Given this conclusion, we then plot the values of
$\gamma_1'$ as a function of the distance from $u=1$ and extrapolate to zero
distance.  This yields the value 
\beq
\gamma_1' = 2.50 \pm 0.01
\label{gamma1ser}
\eeq
This strongly supports the following inference that we shall make for the
exact value of this exponent:
\beq
\gamma_1' = \frac{5}{2}
\label{gamma1}
\eeq
Hence, in particular, 
\beq
\gamma_1'=2\gamma_e'
\label{gammae1}
\eeq
We note that the relation (\ref{gammae1}) can be understood if one
re-expresses $\bar\chi$ as a function of the elliptic modulus variable
$k_<$ in eq. (\ref{kl}), since $k_<$ diverges at $u=1$ with an exponent which
is twice as large as the exponent describing its divergence at $u=u_e$, i.e.,
$k_< \sim (1-u)^{-1}$ as $ u \to 1$, while 
$k_< \sim (1-u/u_e)^{-1/2}$ as $u \to u_e$. 

\subsection{Singularity at $u=-1$}

    We have studied the singularity in $\bar\chi$ at $u=-1$ (denoted $u_s$)
by using the series for $\bar\chi$ in the variable $u$ and also transforming
this series to one in the elliptic modulus variable $k_<$.  The series in $k_<$
showed a greater sensitivity to this singularity, and therefore we concentrate
on the results from our analysis of this series.  The reason for this greater
sensitivity is clear; the series in $u$ is strongly affected by the fact that,
as one can see from Fig. 2(a), there is an intervening singular line segment
protruding into the FM phase and ending at $u=u_e$, in front of the point
$u=-1$ as one moves out from the origin along the negative $Re(u)$ axis.  The
transformation from $u$ to $k_<$ maps the singular endpoint at $u=u_e$ away to
$-\infty$ and maps the singularity at $u=1$ to $\infty$ in the $k_<$ plane,
thereby leaving the singularity at $u=-1$ as the nearest to the origin. 
Specifically, the image of the singular line segment from $u=u_e$ leftward
to $u=-1$ is the semi-infinite line segment from $-\infty$ to $-1$ in the
$k_<$ plane.  The line segment from $u=1/u_e$ to $u=-1$ has the same image,
again the segment from $-\infty$ to $-1$ in the $k_<$ plane, while the unit
circle $|u|=1$ maps to the line segment from 1 to $\infty$ in this plane. 
The series in $k_<$ has the form 
$\bar\chi = (1/4)(k_<)^2\sum_{n=0}^{\infty} c_n'(k_<)^n$, and, as 
before, we actually analyse the reduced function 
$\bar\chi_r = 4(k_<)^{-2}\bar\chi$.  
Using the Taylor series expansion of $k_<$ near $u=-1$, 
\beq
k_< = -1 - 2^{-5}(1+u)^4 + O((1+u)^5)
\label{klnearm1}
\eeq
it follows that as $k_< \to -1$ and $u \to -1$, 
the singular form $\bar\chi \sim (1+k_<)^{-\gamma_{s,k_<}'}$ 
corresponds to $\bar\chi \sim (1+u)^{-\gamma_{s}'}$, with 
\beq
\gamma_s' = 4\gamma_{s,k_<}'
\label{gammasrel}
\eeq
Our results from the differential approximants for this series are 
given in Table 6 and its continuation.  Because
the actual pole positions have small imaginary parts, typically a few times
$10^{-4}$ of the size of the real part, there are resultant imaginary
parts in the values of the corresponding exponent $\gamma_s'$ from the
differential approximants.  Since the exact singularity in $\bar\chi(u)$ 
is at the real value $u=-1$, and since $\bar\chi(u,h=i\pi/2)$ is real for 
real $u$, we know that $\gamma_s'$ at $u=(k_<)=-1$ is real.  Given this and the
relation (\ref{gammasrel}), it follows that we may take only the real parts of
the exponents from the differential approximants to the series in $k_<$, and
we do so. 

\begin{table}
\begin{center}
\begin{tabular}{|c|c|c|c|} \hline \hline  & & & \\
$[L/M_0;M_1]$ & $(k_<)_{sing}$ & $|(k_<)_{sing}+1|$ & $Re(\gamma_{s,k_<}')$ \\
 & & & \\
\hline\hline
$[3/6;7]$ & $-0.9992689 - 0.0002556 i$ & $7.7 \times 10^{-4} $ & $  0.2195 $ 
\\ \hline
$[3/7;6]$ & $-0.9992875 - 0.0002801 i$ & $7.7 \times 10^{-4}$ & $  0.2204 $ 
\\ \hline
$[3/7;8]$ & $-0.9992627 + 0.0004390 i$ & $8.6 \times 10^{-4}$ & $  0.2230 $ 
\\ \hline
$[3/7;9]$ & $-0.9993572 + 0.0003351 i$ & $7.2 \times 10^{-4}$ & $  0.2283 $ 
\\ \hline
$[3/8;7]$ & $-0.9992612 + 0.0004332 i$ & $8.6 \times 10^{-4}$ & $  0.2228 $ 
\\ \hline
$[3/8;8]$ & $-0.9991777 + 0.0000793 i$ & $8.3 \times 10^{-4}$ & $  0.2147 $ 
\\ \hline
$[3/9;7]$ & $-0.9993502 + 0.0003284 i$ & $7.3 \times 10^{-4} $ & $  0.2277 $ 
\\ \hline
$[4/5;7]$ & $-0.9992779 - 0.0001344 i$ & $7.3 \times 10^{-4} $ & $  0.2205 $ 
\\ \hline
$[4/7;5]$ & $-0.9992970 - 0.0001857 i$ & $7.3 \times 10^{-4} $ & $  0.2213 $ 
\\ \hline
$[4/7;8]$ & $-0.9992847 + 0.0002823 i$ & $7.7 \times 10^{-4} $ & $  0.2232 $ 
\\ \hline
$[4/8;7]$ & $-0.9992808 + 0.0002780 i$ & $7.7 \times 10^{-4} $ & $  0.2229 $ 
\\ \hline
$[5/6;8]$ & $-0.9992436 + 0.0002910 i$ & $8.1 \times 10^{-4} $ & $  0.2207 $ 
\\ \hline
$[5/7;7]$ & $-0.9994218 + 0.0003067 i$ & $6.5 \times 10^{-4} $ & $  0.2319 $ 
\\ \hline
$[5/8;6]$ & $-0.9992426 + 0.0002845 i$ & $8.1 \times 10^{-4} $ & $  0.2206 $ 
\\ \hline
$[6/5;5]$ & $-0.9990412 + 0.0000501 i$ & $9.6 \times 10^{-4} $ & $  0.2087 $ 
\\ \hline
$[6/6;6]$ & $-0.9991672 + 0.0005305 i$ & $9.9 \times 10^{-4} $ & $  0.2178 $ 
\\ \hline
$[6/6;7]$ & $-0.9994966 + 0.0002217 i$ & $5.5 \times 10^{-4} $ & $  0.2368 $ 
\\ \hline
$[6/7;6]$ & $-0.9994911 + 0.0002243 i$ & $5.6 \times 10^{-4} $ & $  0.2364 $ 
\\ \hline
\end{tabular}
\end{center}
\caption{Values of pole $(k_<)_{sing}$ near $k_<=-1$, i.e. $u=-1$ 
(denoted $u_s$), distance from this point, $|(k_<)_{sing}+1|$, and 
real part of exponent, $Re(\gamma_{s,k_<}')$ from differential approximants to 
low-temperature series for $\bar\chi_r(k_<,h=i\pi/2)$.  We list 
only the approximants which satisfy the accuracy criterion
$|(k_<)_{sing}+1| < 1 \times 10^{-3}$.}
\label{table5}
\end{table}

\begin{table}
\begin{center}
\begin{tabular}{|c|c|c|c|} \hline \hline
$[7/4;6]$ & $-1.0004373 + 0.0006097 i$ & $7.5 \times 10^{-4} $ & $  0.2916 $ 
\\ \hline
$[7/5;6]$ & $-0.9993254 + 0.0004203 i$ & $7.9 \times 10^{-4} $ & $  0.2269 $ 
\\ \hline
$[7/5;7]$ & $-0.9993665 + 0.0002981 i$ & $7.0 \times 10^{-4} $ & $  0.2286 $ 
\\ \hline
$[7/6;4]$ & $-0.9998382 + 0.0005474 i$ & $5.7 \times 10^{-4} $ & $  0.2573 $ 
\\ \hline
$[7/6;5]$ & $-0.9993332 + 0.0004066 i$ & $7.8 \times 10^{-4} $ & $  0.2273 $ 
\\ \hline
$[7/6;6]$ & $-0.9993588 + 0.0003212 i$ & $7.2 \times 10^{-4} $ & $  0.2285 $ 
\\ \hline
$[7/7;5]$ & $-0.9993650 + 0.0003025 i$ & $7.0 \times 10^{-4} $ & $  0.2286 $ 
\\ \hline
$[8/4;6]$ & $-0.9993815 + 0.0001580 i$ & $6.4 \times 10^{-4} $ & $  0.2286 $ 
\\ \hline
$[8/5;6]$ & $-0.9993680 + 0.0002999 i$ & $7.0 \times 10^{-4} $ & $  0.2287 $ 
\\ \hline
$[8/6;4]$ & $-0.9993746 + 0.0002485 i$ & $6.7 \times 10^{-4} $ & $  0.2288 $ 
\\ \hline
$[8/6;5]$ & $-0.9993660 + 0.0003035 i$ & $7.0 \times 10^{-4} $ & $  0.2287 $ 
\\ \hline
$[9/4;5]$ & $-0.9992599 + 0.0000684 i$ & $7.4 \times 10^{-4} $ & $  0.2199 $ 
\\ \hline
$[9/4;6]$ & $-1.0005762 + 0.0005457 i$ & $7.9 \times 10^{-4} $ & $  0.3106 $ 
\\ \hline
$[9/5;4]$ & $-0.9992690 + 0.0001240 i$ & $7.4 \times 10^{-4} $ & $  0.2209 $ 
\\ \hline
$[9/5;5]$ & $-0.9992719 + 0.0002258 i$ & $7.6 \times 10^{-4} $ & $  0.2218 $ 
\\ \hline
$[9/6;4]$ & $-0.9995104 + 0.0003921 i$ & $6.3 \times 10^{-4} $ & $  0.2390 $ 
\\ \hline
$[10/4;4]$ & $-0.9994550 - 0.0001668 i$ & $5.7 \times 10^{-4} $ & $ 0.2311 $ 
\\ \hline
$[10/4;5]$ & $-0.9991991 + 0.0001605 i$ & $8.2 \times 10^{-4} $ & $ 0.2164 $ 
\\ \hline
$[10/5;4]$ & $-0.9992330 + 0.0001856 i$ & $7.9 \times 10^{-4} $ & $ 0.2189 $ 
\\ \hline
\hline
\end{tabular}
\end{center}
\caption{Continuation of table of differential approximants for $\bar\chi_r$
series in the variable $k_<$ near $u=k_<=-1$.}
\label{table5a}
\end{table}
From this study, we obtain for the position of the singularity
\beq
(k_<)_{sing} = -0.9993 \pm 0.0001
\label{klposition}
\eeq
consistent with the expectation $(k_<)_{sing}=-1$, or equivalently, 
$u_{sing}=u_s=-1$.  As is
evident, the values of $Re(\gamma_s')$ are almost all slightly below 0.25;
however, when we carry out our method of plotting the values as a function 
of the distance $|(k_<)+1|$ and extrapolating to zero distance from the exact 
singularity, the extrapolated value is actually slightly above 0.25.
Accordingly, we give a conservative estimate
\beq
\gamma_{s,k<}'=0.250 \pm 0.020
\label{gammaskl}
\eeq
and hence, using (\ref{gammasrel}), 
\beq
\gamma_s' = 1.00 \pm 0.08
\label{gammasser}
\eeq
This supports the conclusion, which we shall draw, that the exact value of 
this exponent is 
\beq
\gamma_s'=1
\label{gammas}
\eeq
We show our summary of exponents in table 8. The exponent relation 
$\alpha_{u,ph.}' + 2\beta_u' + \gamma_{u,ph.}'=2$ is evidently satisfied at 
all three of the singularities $u=u_e$, $u=1$, and $u=-1$. 

\begin{table}
\begin{center}
\begin{tabular}{|c|c|c|c|c|c|} \hline \hline & & & & & \\ 
$u$ & $\alpha_{u,FM}'$ & $\alpha_{u,AFM}'$ & $\beta_u$ 
& $\gamma_{u}'$ & $\alpha_{u,ph.}' + 2\beta_u' + \gamma_{u,ph.}'$ \\
& & & & & \\
\hline \hline
$u_e=-(3-2^{3/2})$ & 1 & $-$ & $-1/8$ & $5/4$ & 2 \\ \hline
1 & 0 finite$^*$ & 0 finite$^*$ & $-1/4$ & $5/2$ & 2 \\ \hline
$u_s=-1$ & 0 finite & 0 finite & $1/2$ & $1$ & 2 \\ \hline
\hline
\end{tabular}
\end{center}
\caption{Exponents at singularities in the 2D Ising model with $h=\pm i \pi/2$.
The results for $\alpha_u'$ and $\beta_u$ are exact; the results for
$\gamma_u'$ are our conclusions for the exact values from our series analysis.
The notation $-$ indicates that the point cannot be approached from within the
given phase.  For the singularity of $C$ at $u=1$ (marked with a $^*$), the
values of $\alpha'$ correspond to evaluating $K=-(1/4)\ln(1)=0$ on the 
principal Reimann sheet of the logarithm, as discussed in the text.}
\label{table6}
\end{table}

\section{Extraction and Analysis of Low-Temperature Series for
$\bar\chi^{(a)}$}

   We have also investigated the complex-temperature singularities in the 
staggered susceptibility $\bar\chi^{(a)}$ for the present model.  To do this,
we have extracted and analysed the low-temperature series expansion for this
function, using the low-temperature, high staggered field series expansions for
the free energy of the Ising model on the square lattice calculated by the
King's College group \cite{sykes1,tlow1}.  These are denoted antiferromagnetic
polynomials in these papers and were calculated to order $O(w^{11})$ in
Ref. \cite{tlow1}, where, as before, $w=1/u$ is the low-temperature expansion
variable in the AFM phase. 
The antiferromagnetic polynomials were apparently not 
calculated to higher order subsequently \cite{gaunt}.  
We have extracted from these the resultant 
low-temperature series expansion for $\bar\chi^{(a)}$ for $h=i\pi/2$, which is 
\beqs
\bar\chi^{(a)}(h=i\pi/2) & = 4w^2 \Big[ -1-8w^2+24w^3-135w^4+648w^5-3336w^6 
\nonumber \\
& +17240w^7-90501w^8+479192w^9 + O(w^{10}) \Bigr ]
\label{chiah}
\eeqs
For reference, we recall that the series for $\bar\chi^{(a)}$ for 
 $h=0$ on this lattice is \cite{sykes1,tlow1}
\beq
\bar\chi^{(a)}(h=0) = 4w^2 \Big[ 1+4w^2+8w^3+39w^4+152w^5+672w^6
+3016w^7+13989w^8+66664w^9 + O(w^{10}) \Bigr ]
\label{chia0}
\eeq
The series (\ref{chiah}) is much shorter than the one which we extracted for
$\bar\chi$, given by eq. (\ref{chihuseries}) and Table 1, and hence one does
not expect to derive results for $\bar\chi^{(a)}$ which are as precise as those
which we obtained for $\bar\chi$.  As before,
we have used both dlog Pad\'{e} and differential approximants for this
analysis.  

    We study first the vicinity of the singular point $w=1$.  
In Table 9 we list the pole locations and resultant exponents for
the approximants which satisfy the accuracy requirement $|w_{sing}-1| \le
0.05$.  One would not normally expect
a dlog Pad\'{e} approximant of such low order as [1/2] to yield an accurate
result; however, it happens that the denominator of this approximant is 
$\propto (1-w)(1-(11/2)w)$, so that it produces a location for the pole which 
is exact.  For this reason, it yields a much better determination of the
associated exponent than would otherwise have been the case. 
This fortuitously accurate approximant, and the best differential approximant,
$[3/2;2]$, both give $\gamma_{1,a}'$ values of about 2.5.  From the full set,
we infer the crude result
\beq
\gamma_{1,a}'=2.5 \pm 0.5
\label{gamma0a}
\eeq
This is consistent with the exact value $\gamma_{1,a}'=5/2$ and hence with the 
equality $\gamma_{1,a}'=\gamma_1'$.  However, clearly the results for
$\gamma_{1,a}'$ are much less precise than our determination of $\gamma_1'$.

\begin{table}
\begin{center}
\begin{tabular}{|c|c|c|c|} \hline \hline & & & \\ 
PA or DA & $w_{sing}$ & $|w_{sing}-1|/$ & $\gamma_{1,a}'$ \\
& & & \\
\hline \hline
$[1/2]$ & $1$        & $0$                   & $2.462$ \\ \hline
$[2/3]$ & $1.04562$  & $4.6 \times 10^{-2} $ & $2.626$ \\ \hline
$[2/4]$ & $0.950447$ & $5.0 \times 10^{-2} $ & $2.063$ \\ \hline
$[4/4]$ & $0.956193$ & $4.4 \times 10^{-2} $ & $2.063$ \\ \hline
$[3/5]$ & $0.969809$ & $3.0 \times 10^{-2} $ & $2.162$ \\ \hline
$[0/3;4]$ & $0.955169$ & $4.5 \times 10^{-2} $ & $2.057$ \\ \hline
$[0/4;2]$ & $0.955792$ & $4.4 \times 10^{-2} $ & $2.172$ \\ \hline
$[2/2;2]$ & $0.999046$ & $0.95 \times 10^{-3} $ & $2.554$ \\ \hline
$[3/2;2]$ & $0.952282$ & $4.8  \times 10^{-2} $ & $2.357$ \\ \hline
\hline
\end{tabular}
\end{center}
\caption{Values of pole near $w=1$ (denoted $w_1$), normalised 
distance from this point, $|w_{sing}-1|$, and exponent $\gamma_{1,a}'$ 
from dlog Pad\'{e} and differential approximants to low-temperature series for 
$\bar\chi^{(a)}_r$.  
We list only the Pad\'{e} and differential approximants which
satisfy the accuracy criterion $|w_{sing}-1| \le 0.05$.}
\label{table7}
\end{table}

   We also studied the series in the vicinity of the singular endpoint 
$w=w_e=-(3-2^{3/2})$ (i.e., $u=u_{oe}=1/u_e=-(3+2^{3/2})$).  To optimise the 
sensitivity, we calculated and analysed series in 
transformed variables to map the singularity at $w=1$ away.  We 
required these variables to be equal to $w$ for small $w$ and to map $w=\pm
\infty$ to $\pm \infty$, respectively.  Two such variables were 
$w' = w(1+w/8)(1-w)^{-1}$ and $w'' = w(1-w)^{-1}\sinh w$.  The series in the
transformed variables did slightly better in locating the pole positions in
$w_e'$ and $w_e''$ corresponding to $w_e$.  The dlog Pad\'{e} and differential
approximants indicated that $\bar\chi^{(a)}$ has a divergent singularity at
$w_e$ and yielded values for the associated exponent $\gamma_{oe,a}'$ in the 
range from about $0.2$ to $0.4$. Given our exact results $\alpha_{oe}'=1$ for
the specific heat and $\beta_{oe}=-1/4$ for the staggered magnetisation, a 
value within the above range for $\gamma_{oe,a}'$ would indicate a violation of
the exponent relation $\alpha_{oe}'+2\beta_{oe}+\gamma_{oe}'=2$. In this 
context, it is of interest 
to note that we have already found violations of the relation 
$\alpha + 2\beta + \gamma=2$ at complex-temperature singularities, e.g., in 
the zero-field Ising model on the square lattice at $u=u_s=-1$, as 
approached from within the PM phase, where 
$\alpha_s=0$, $\beta_s=1/4$, and $\gamma_s < 0$ (since $\bar\chi$ has a finite
non-analyticity for the approach from within the PM phase) \cite{chisq}, and in
the zero-field Ising model on the honeycomb lattice, at the point 
$z=z_\ell=-1$, as approached from within the FM phase, where 
$\alpha_{\ell}'=2$, $\beta_{\ell}=-1/4$, and $\gamma_{\ell}'=5/2$, so that 
$\alpha_{\ell}'+2\beta_\ell+\gamma_{\ell}'=4$ \cite{chihc}.

\section{Complex-Temperature Behaviour of the Correlation Length}
\label{xisection}

   In this section we shall study the complex-temperature behaviour of the
correlation length.  To do this, we make use of a calculation of the asymptotic
form of the spin-spin correlation function along a row (or equivalently, 
column), $<\sigma_{0,0}\sigma_{n,0}>$, for large $n$ \cite{mw} (where, without
loss of generality, one may take $n > 0$.  From this calculation, carrying 
out an analytic continuation to complex temperature, we obtain, for $n \to 
\infty$, 
\begin{displaymath}
<\sigma_{0,0}\sigma_{n,0}>_{conn.} \sim  -(2/\pi) (1-u^2)^{-1}M^2 n^{-1} 
u (-u)^n
\end{displaymath}
\beq
 = -(2/\pi)(1-u)^{-3/2}(1+6u+u^2)^{-1/4}n^{-1} u (-u)^n
\label{asymcorfun}
\eeq
where we have used the exact expression for $M$, (\ref{mh}). This analytic
continuation applies within the FM phase.  Extracting the correlation 
length $\xi$ in the usual way as $\xi^{-1} = -\lim_{r \to \infty} r^{-1}
\ln(<\sigma_{\bf 0}\sigma_{\bf r}>_{conn.})$, where 
$r \equiv |{\bf r}|$, we find
\beq
\xi_{row}^{-1} = -\ln(-u)
\label{xirow}
\eeq
For usual physical second-order critical points, one can use the connected
2-spin correlation function for any ${\bf r}$, with $|{\bf r}| \to \infty$, to
extract the correlation length $\xi$.  However, in our previous work
\cite{chisq,chitri}, we found that at the complex-temperature singular point 
$u=u_s=-1$ in the zero-field Ising model on the square lattice, the 
correlation length defined from the diagonal connected 2-spin correlation 
function diverges with a different exponent, $\nu_{s,diag}'=2$, than the 
exponent $\nu_{s}'=1$ describing the divergence in the correlation length 
defined from off-diagonal (e.g., row) correlation functions. 
In view of this, we include the suffix $row$ in (\ref{xirow}) for clarity.  
We now consider three particular singular points which can be
approached from within the complex-temperature FM phase, viz., $u=u_e$, $u=-1$,
and $u=1$.  As $u \to u_e$, the 2-spin correlation function 
(\ref{asymcorfun}) diverges, like $(1-u/u_e)^{-1/4}$, 
because of the divergence in the prefactor $M^2$, but the 
correlation length $\xi_{row}$ remains finite, with 
$\xi^{-1}_{row}=-\ln(-u_e)=1.7627...$ at $u=u_e$.  If this feature of a finite
correlation length applied to all of the connected 2-spin correlation 
functions, precisely at $u=u_e$ as well as for points approaching $u_e$ from
within the complex-temperature FM phase, then by the same argument as was used
in Ref. \cite{ms}, it would follow that the only singularity in 
$\bar\chi$ would arise from the divergent $M^2$
prefactor.\footnote{Define $<\sigma_{\bf 0}\sigma_{\bf r}> \ =
M^2c({\bf r})$.  Then $\bar\chi = M^2 \sum_{\bf r} c({\bf r})$.  For purposes 
of analysing divergences, the asymptotic behaviour of the sum can be 
approximated by that of the integral $\int d^2r \ c(\bf r)$ for large $r$ (with
a short-distance cutoff on the latter).  If
$c({\bf r}) \sim r^{-p} e^{-r/\xi}$ as $r \to \infty$, this integral is
finite. Therefore, a divergence in $\bar\chi$ would arise solely from a
divergence in the prefactor $M^2$.  This was noted in Ref. \cite{chitri}.} 
We know, however, that the above premise cannot be true, since then the 
susceptibility
exponent at $u_e$ (as approached from within the FM phase) would be 1/4,
whereas we found that $\gamma_e'=5/4$.  The fact that the susceptibility
diverges with an exponent different from that arising from the divergent $M^2$
prefactor shows that at least some connected 2-spin correlation functions 
must decay like a power law, i.e. the associated correlation length must be
divergent, at $u=u_e$.  To get more information on this, it would be useful 
to carry out analytic calculations of the asymptotic forms of the general 
2-spin correlation functions $<\sigma_{0,0}\sigma_{m,n}>$ in the present 
model, near to and at this singular point. 

  As $u \to 1$, the 2-spin correlation function (\ref{asymcorfun})
again diverges, like $(1-u)^{-3/2}$, and the correlation length 
$\xi_{row}$ is finite: $\xi_{row}^{-1}=-\ln(-1) = -i \pi$ (for the principal 
Riemann sheet of the logarithm).  This is a case similar that discussed in
\cite{ms} where $Re(\xi^{-1})=0$ but $Im(\xi^{-1}) \ne 0$.  

As $u \to -1$, each 2-spin correlation function is finite, but the correlation
length does diverge, with exponent
\beq
\nu_s' = 1
\label{nus}
\eeq
If one were to use the exponent relation $\gamma_s'=\nu_s'(2-\eta_s)$, then
with our inference $\gamma_s'=1$ in eq. (\ref{gammas}), it would follow that
$\eta_s=1$.  However, we have shown previously \cite{chisq} that one must use
caution in trying to apply such exponent relations at complex-temperature
singularities, since different connected spin-spin correlation functions may be
characterised by correlation lengths which diverge with different exponents
$\nu$. 

   This type of analysis can also be done with the staggered 2-spin 
correlation functions (c.f. (\ref{sigmatilde})) 
\beq
<\tilde \sigma_{0,0}\tilde \sigma_{n,0}>_{conn.} \ = (-1)^n
<\sigma_{0,0}\sigma_{n,0}>_{conn.}
\label{sstag}
\eeq
In particular, as $u \to 1/u_e$ (i.e., $w \to w_e$), these correlation
functions diverge, like $(1-w/w_e)^{-1/4}$, because of the divergent prefactor
$M_{st}^2$.  However, the correlation length $\xi_{row,AFM}$ remains finite.
If this behaviour characterised all of the staggered 2-spin correlation
functions, at $w_e$ as well as in the vicinity of $w_e$, then the only
divergence in $\bar\chi^{(a)}$ would arise from the $M_{st}^2$ prefactor, and
hence $\gamma_{e,a}'=1/4$.  This value is consistent with our results from 
the analysis of the low-temperature series for $\bar\chi^{(a)}$.  This merits
further study. 

\section{Exact Solution at $u=1$ for Arbitrary $H$}

   In the body of this paper, we have investigated singularities in the square
lattice Ising model as functions of complex-temperature, for the fixed value 
of external magnetic field, $h=i\pi/2$ (or $h=-i\pi/2$).  It is also of 
interest to study the complementary problem of singularities as a function 
of $h$ for fixed $K$ or $u$.  Indeed, in pursuing such a study, Yang and Lee 
were led to their celebrated circle theorem on the zeros of the partition 
function for the Ising model in the complex $e^{2h}$ plane \cite{ly,yl}.  
Here, we would like to mention some elementary results which elucidate how
various quantities become singular at a particularly simple point, $u=1$, as
$h$ is varied.  These results may be combined with our determination of the
exact singularities in $f$, $U$, $C$, and $M$ as one approaches this point 
by varying $u$.  
At $J=0$, hence $K=0$ and $u=1$, the partition function reduces to a 
single-site problem, which can easily be calculated exactly for arbitrary $H$,
dimensionality, and lattice type. 
We find, independent of the dimensionality and lattice type, 
\beq
f(u=1,h) = \ln(2 \cosh h)
\label{fu1}
\eeq
\beq
U = -H \tanh h
\label{hu1}
\eeq
\beq
k_B^{-1} C = \frac{h^2}{\cosh^2 h}
\label{cu1}
\eeq
\beq
M = \tanh h
\label{mu1}
\eeq
\beq
\bar\chi = \frac{1}{\cosh^2 h}
\label{chiu1}
\eeq
Further, the $m$-point correlation functions factorise trivially and are
independent of the positions of the spins:
\beq
<\sigma_{\bf r_1} \cdot \cdot \cdot \sigma_{\bf r_m}> \ = \ 
<\sigma_{\bf r_1}>^m \ = M^m = (\tanh h)^m
\label{corfunu1}
\eeq
To study the singularities of these functions, we must define new exponents,
since the usual critical exponents apply to singularities of thermodynamic
quantities as functions of $T$.  To avoid a profusion of new symbols, we
shall use the same Greek letters as for the respective $T$-dependent
singularities in thermodynamic quantities, but use a superscript $(h)$ to
indicate that they describe the singularity as a function of $h$ for $K=0$.  
Thus, for the leading singularity in the specific heat, as
a function of $h$, at the point $h=h_s$ ($s$ denotes a generic singularity
here) for fixed $K=0$ (hence $u=1$), we shall write
\beq
C(h)_{sing.} \sim A_{C,s,dir.}^{(h)}(h-h_s)^{-\alpha_{s,dir.}^{(h)}} 
\label{csing.}
\eeq
where $dir.$ denotes the direction, in the complex $h$ plane, from which one
approaches the singular point $h_s$.  
Similarly, we shall write
\beq
\chi_{sing.} \sim A_{\chi,s,dir.}^{(h)}(h-h_s)^{\gamma_{s,dir.}^{(h)}}
\label{chising}
\eeq
and so forth for the singularities in other quantities.  From (\ref{cu1}), 
it is evident that $C$ diverges for 
\beq
h = (2n+1)\frac{i \pi}{2} \ , \qquad n \in Z
\label{hcdiv}
\eeq
with corresponding exponent
\beq
\alpha_1^{(h)}=2
\label{alphau1}
\eeq
for any direction of approach to any of the singular points (\ref{hcdiv}). 
The internal energy itself also diverges at these points, with the 
exponent $\alpha_{1}^{(h)}-1=1$ and vanishes at the set 
\beq
h = n i \pi \ , \qquad n \in Z
\label{huzero}
\eeq
The magnetisation vanishes and diverges at the same set of points as the
internal energy $U$ (c.f. eqs. (\ref{huzero}) and (\ref{hcdiv}) with the
respective exponents 
\beq
\beta_{1,zero}^{(h)}=1
\label{betau1zero}
\eeq
and 
\beq
\beta_{1,div}^{(h)}=-1
\label{betau1div}
\eeq
again, independent of the direction of approach to these points in the
complex $h$ plane. 
The susceptibility diverges at the points (\ref{hcdiv}) with exponent
\beq
\gamma_1^{(h)}=2
\label{gammau1}
\eeq

It is useful to evaluate the general results above for the interesting special
case of complex $h=h_r + i \pi/2$ where $h_r$ is real.  For this case, we have 
\beq
f = \ln(2 i \sinh h_r)
\label{fhr}
\eeq
\beq
U = -\frac{H}{\tanh h_r}
\label{uhr}
\eeq
\beq
k_B^{-1} C = -\frac{(h_r+i\pi/2)^2}{\sinh^2 h_r}
\label{chr}
\eeq
\beq
M = \frac{1}{\tanh h_r}
\label{mhr}
\eeq
\beq
\bar\chi = -\frac{1}{\sinh^2 h_r}
\label{chihr}
\eeq
and
\beq
<\sigma_{\bf r_1} \cdot \cdot \cdot \sigma_{\bf r_m}> \ = (\tanh h_r)^{-m}
\label{corfunu1hr}
\eeq
As $h_r \to 0$ so that $h \to i\pi/2$, we recover our previous results, that 
$U$ and $M$ diverge linearly while $C$ and $\bar\chi$ diverge quadratically. 

  For loose-packed lattices, it is straightforward to extend these results 
to consider a staggered rather than uniform external field, $H_{st}$, i.e. to
consider the partition function 
\beq
Z = \sum_{\{ \sigma_n \} } \exp( \sum_{n}(-1)^{p(n)}h_{st}\sigma_n)
\label{zstag}
\eeq
where $p(n)$ was defined in eq. (\ref{pfun}) and $h_{st}=\beta H_{st}$. 
Since the summations over the spins on the even and odd
sublattices are decoupled, one finds the same equations as before, but with $H$
replaced by $H_{st}$, i.e., $f=\ln(2\cosh h_{st})$, $U = -H_{st}\tanh h_{st}$,
$C=k_B h_{st}^2/\cosh^2 h_{st}$, and $M_{st}=\tanh h_{st}$.  The staggered
susceptibility is $\bar\chi^{(a)}= 1/\cosh^2 h_{st}$.

\section{Conclusions}

   In this paper, we have studied a natural generalisation of an exactly solved
model from real non-negative temperature to complex temperature.  This is the
Ising model on the square lattice in an external magnetic field given by 
$\beta H = \pm i\pi/2$, first solved by Lee and Yang \cite{ly}.  We have 
worked out the complex-temperature phase boundaries, as shown in Fig. 2.  We 
have also extracted a low-temperature series expansion for the susceptibility,
$\bar\chi$.  From an analysis of this series using dlog Pad\'{e} and 
differential approximants, we conclude that $\bar\chi$ has divergent 
singularities at $u=u_e=-(3-2^{3/2})$ with exponent $\gamma_e'=5/4$, at 
$u=1$ with exponent $\gamma_1'=5/2$, and at $u=u_s=-1$ with exponent 
$\gamma_s'=1$.  We have also studied the staggered susceptibility.  
Using exact results, we have determined the
complex-temperature singularities of the specific heat and the uniform and
staggered magnetisation.  We are currently in the process of extending our
studies to other 2D lattices.  The findings show again that even though the
Ising model has a very simple Hamiltonian, it exhibits a fascinating 
richness of properties.

This research was supported in part by the NSF grant PHY-93-09888.  One of us
(R.S.) thanks Prof. C. N. Yang for a discussion of Ref. \cite{ly} and
Profs. D. S. Gaunt and A. J. Guttmann for discussions of the current status of
low-temperature series expansions. 

\vfill 
\eject

\vfill
\eject

\begin{center}
{\Large \bf Appendix} 
\end{center}

   Here we list our differential approximant results relevant for the
singularity in $\bar\chi(u,h=i\pi/2)$ at $u=u_e$. 

\begin{table}
\begin{center}
\begin{tabular}{|c|c|c|c|} \hline \hline \\ 
$[L/M_0;M_1]$ & $u_{sing}$ & $|u_{sing}-u_e|/|u_e|$ & $\gamma_e'$ \\
& & & \\
\hline \hline
$[0/4;6]$ & $-0.1715614$ & $6.7 \times 10^{-5} $ & $   1.2455$ \\ \hline
$[0/5;5]$ & $-0.1715743$ & $8.1 \times 10^{-6} $ & $   1.2474$ \\ \hline
$[0/5;6]$ & $-0.1715716$ & $7.2 \times 10^{-6} $ & $   1.2470$ \\ \hline
$[0/5;7]$ & $-0.1715696$ & $1.9 \times 10^{-5} $ & $   1.2467$ \\ \hline
$[0/6;4]$ & $-0.1715688$ & $2.4 \times 10^{-5} $ & $   1.2466$ \\ \hline
$[0/6;5]$ & $-0.1715717$ & $6.6 \times 10^{-6} $ & $   1.2470$ \\ \hline
$[0/6;7]$ & $-0.1715661$ & $3.9 \times 10^{-5} $ & $   1.2461$ \\ \hline
$[0/6;8]$ & $-0.1715659$ & $4.1 \times 10^{-5} $ & $   1.2460$ \\ \hline
$[0/7;5]$ & $-0.1715706$ & $1.3 \times 10^{-5} $ & $   1.2469$ \\ \hline
$[0/7;6]$ & $-0.1715662$ & $3.9 \times 10^{-5} $ & $   1.2461$ \\ \hline
$[0/7;7]$ & $-0.1715659$ & $4.1 \times 10^{-5} $ & $   1.2460$ \\ \hline
$[0/7;8]$ & $-0.1715660$ & $4.0 \times 10^{-5} $ & $   1.2460$ \\ \hline
$[0/7;9]$ & $-0.1715628$ & $5.8 \times 10^{-5} $ & $   1.2456$ \\ \hline
$[0/8;6]$ & $-0.1715658$ & $4.1 \times 10^{-5} $ & $   1.2460$ \\ \hline
$[0/8;7]$ & $-0.1715660$ & $4.0 \times 10^{-5} $ & $   1.2460$ \\ \hline
$[0/8;8]$ & $-0.1715654$ & $4.3 \times 10^{-5} $ & $   1.2459$ \\ \hline
$[0/8;10]$ & $-0.1715690$ & $2.2 \times 10^{-5} $ & $  1.2469$ \\ \hline
$[0/9;7]$ & $-0.1715599$ & $7.6 \times 10^{-5} $ & $   1.2456$ \\ \hline
$[0/9;8]$ & $-0.1715549$ & $1.0 \times 10^{-4} $ & $   1.2468$ \\ \hline
$[0/9;9]$ & $-0.1715754$ & $1.5 \times 10^{-5} $ & $   1.2505$ \\ \hline
$[0/9;10]$ & $-0.1715716$ & $7.2 \times 10^{-6} $ & $  1.2481$ \\ \hline
$[0/10;8]$ & $-0.1715584$ & $8.5 \times 10^{-5} $ & $  1.2456$ \\ \hline
$[0/10;9]$ & $-0.1715718$ & $6.4 \times 10^{-6} $ & $  1.2482$ \\ \hline
$[1/5;5]$ & $-0.1715621$ & $6.3 \times 10^{-5} $ & $   1.2449$ \\ \hline
$[1/5;6]$ & $-0.1715609$ & $7.0 \times 10^{-5} $ & $   1.2447$ \\ \hline
$[1/5;7]$ & $-0.1715625$ & $6.0 \times 10^{-5} $ & $   1.2450$ \\ \hline
$[1/6;5]$ & $-0.1715609$ & $7.0 \times 10^{-5} $ & $   1.2446$ \\ \hline
$[1/6;6]$ & $-0.1715615$ & $6.7 \times 10^{-5} $ & $   1.2448$ \\ \hline
$[1/6;7]$ & $-0.1715649$ & $4.7 \times 10^{-5} $ & $   1.2457$ \\ \hline
$[1/6;8]$ & $-0.1715676$ & $3.1 \times 10^{-5} $ & $   1.2466$ \\ \hline 
$[1/7;5]$ & $-0.1715624$ & $6.1 \times 10^{-5} $ & $   1.2450$ \\ \hline
$[1/7;6]$ & $-0.1715647$ & $4.8 \times 10^{-5} $ & $   1.2457$ \\ \hline
\hline
\end{tabular}
\end{center}
\caption{Values of pole near $u_e=-(3-2^{3/2})=-0.171572876...$, normalised
distance from this point, $|u_{sing}-u_e|/|u_e|$, and exponent $\gamma_e'$ 
from differential approximants (DA's) 
to low-temperature series for $\bar\chi_r$.  We list 
only the differential approximants which satisfy the accuracy criterion
$|u_{sing}-u_e|/|u_e| \le 1 \times 10^{-4}$.}
\label{table8a}
\end{table}

\begin{table}
\begin{center}
\begin{tabular}{|c|c|c|c|} \hline \hline
$[1/7;7]$ & $-0.1715669$ & $3.5 \times 10^{-5} $ & $   1.2463$ \\ \hline
$[1/7;8]$ & $-0.1715682$ & $2.7 \times 10^{-5} $ & $   1.2468$ \\ \hline
$[1/7;9]$ & $-0.1715702$ & $1.6 \times 10^{-5} $ & $   1.2475$ \\ \hline
$[1/8;6]$ & $-0.1715676$ & $3.1 \times 10^{-5} $ & $   1.2466$ \\ \hline
$[1/8;7]$ & $-0.1715682$ & $2.7 \times 10^{-5} $ & $   1.2468$ \\ \hline
$[1/8;8]$ & $-0.1715690$ & $2.2 \times 10^{-5} $ & $   1.2471$ \\ \hline
$[1/8;9]$ & $-0.1715700$ & $1.7 \times 10^{-5} $ & $   1.2474$ \\ \hline
$[1/8;10]$ & $-0.1715701$ & $1.6 \times 10^{-5} $ & $  1.2475$ \\ \hline
$[1/9;7]$ & $-0.1715769$ & $2.4 \times 10^{-5} $ & $   1.2513$ \\ \hline
$[1/9;8]$ & $-0.1715699$ & $1.7 \times 10^{-5} $ & $   1.2474$ \\ \hline
$[1/9;9]$ & $-0.1715704$ & $1.4 \times 10^{-5} $ & $   1.2476$ \\ \hline
$[1/10;8]$ & $-0.1715702$ & $1.5 \times 10^{-5} $ & $  1.2475$ \\ \hline
$[2/4;4]$ & $-0.1715557$ & $1.0 \times 10^{-4} $ & $   1.2435$ \\ \hline
$[2/4;5]$ & $-0.1715555$ & $1.0 \times 10^{-4} $ & $   1.2434$ \\ \hline
$[2/4;6]$ & $-0.1715609$ & $7.0 \times 10^{-5} $ & $   1.2446$ \\ \hline
$[2/5;4]$ & $-0.1715555$ & $1.0 \times 10^{-4} $ & $   1.2434$ \\ \hline
$[2/5;5]$ & $-0.1715609$ & $7.0 \times 10^{-5} $ & $   1.2446$ \\ \hline
$[2/5;6]$ & $-0.1715612$ & $6.8 \times 10^{-5} $ & $   1.2445$ \\ \hline
$[2/5;7]$ & $-0.1715651$ & $4.5 \times 10^{-5} $ & $   1.2460$ \\ \hline
$[2/6;4]$ & $-0.1715609$ & $7.0 \times 10^{-5} $ & $   1.2446$ \\ \hline
$[2/6;5]$ & $-0.1715612$ & $6.8 \times 10^{-5} $ & $   1.2446$ \\ \hline
$[2/6;6]$ & $-0.1715671$ & $3.4 \times 10^{-5} $ & $   1.2479$ \\ \hline
$[2/6;7]$ & $-0.1715658$ & $4.2 \times 10^{-5} $ & $   1.2466$ \\ \hline
$[2/6;8]$ & $-0.1715686$ & $2.5 \times 10^{-5} $ & $   1.2466$ \\ \hline
$[2/7;5]$ & $-0.1715649$ & $4.6 \times 10^{-5} $ & $   1.2459$ \\ \hline
$[2/7;6]$ & $-0.1715657$ & $4.2 \times 10^{-5} $ & $   1.2467$ \\ \hline
$[2/7;7]$ & $-0.1715701$ & $1.6 \times 10^{-5} $ & $   1.2463$ \\ \hline
$[2/7;8]$ & $-0.1715688$ & $2.4 \times 10^{-5} $ & $   1.2465$ \\ \hline
$[2/7;9]$ & $-0.1715698$ & $1.8 \times 10^{-5} $ & $   1.2473$ \\ \hline
$[2/8;6]$ & $-0.1715685$ & $2.6 \times 10^{-5} $ & $   1.2466$ \\ \hline
$[2/8;7]$ & $-0.1715688$ & $2.4 \times 10^{-5} $ & $   1.2465$ \\ \hline
$[2/8;8]$ & $-0.1715686$ & $2.5 \times 10^{-5} $ & $   1.2459$ \\ \hline
$[2/8;9]$ & $-0.1715737$ & $4.5 \times 10^{-6} $ & $   1.2501$ \\ \hline
$[2/9;7]$ & $-0.1715693$ & $2.1 \times 10^{-5} $ & $   1.2468$ \\ \hline
$[2/9;8]$ & $-0.1715785$ & $3.3 \times 10^{-5} $ & $   1.2506$ \\ \hline 
$[3/4;4]$ & $-0.1715556$ & $1.0 \times 10^{-4} $ & $   1.2434$ \\ \hline
$[3/4;5]$ & $-0.1715619$ & $6.4 \times 10^{-5} $ & $   1.2449$ \\ \hline
$[3/4;6]$ & $-0.1715632$ & $5.6 \times 10^{-5} $ & $   1.2452$ \\ \hline
\hline
\end{tabular}
\end{center}
\caption{Continuation of DA Table for $u_e$ Singularity.}
\label{table8b}
\end{table}

\begin{table}
\begin{center}
\begin{tabular}{|c|c|c|c|} \hline \hline
$[3/5;3]$ & $-0.1715648$ & $4.7 \times 10^{-5} $ & $   1.2455$ \\ \hline
$[3/5;4]$ & $-0.1715616$ & $6.6 \times 10^{-5} $ & $   1.2448$ \\ \hline
$[3/5;5]$ & $-0.1715628$ & $5.8 \times 10^{-5} $ & $   1.2451$ \\ \hline
$[3/5;6]$ & $-0.1715657$ & $4.2 \times 10^{-5} $ & $   1.2460$ \\ \hline
$[3/5;7]$ & $-0.1715672$ & $3.3 \times 10^{-5} $ & $   1.2463$ \\ \hline
$[3/6;4]$ & $-0.1715637$ & $5.3 \times 10^{-5} $ & $   1.2454$ \\ \hline
$[3/6;5]$ & $-0.1715656$ & $4.2 \times 10^{-5} $ & $   1.2460$ \\ \hline
$[3/6;6]$ & $-0.1715660$ & $4.0 \times 10^{-5} $ & $   1.2466$ \\ \hline
$[3/6;7]$ & $-0.1715689$ & $2.3 \times 10^{-5} $ & $   1.2463$ \\ \hline
$[3/6;8]$ & $-0.1715687$ & $2.4 \times 10^{-5} $ & $   1.2464$ \\ \hline
$[3/7;5]$ & $-0.1715672$ & $3.3 \times 10^{-5} $ & $   1.2463$ \\ \hline
$[3/7;6]$ & $-0.1715690$ & $2.3 \times 10^{-5} $ & $   1.2463$ \\ \hline
$[3/7;7]$ & $-0.1715686$ & $2.5 \times 10^{-5} $ & $   1.2463$ \\ \hline
$[3/7;8]$ & $-0.1715699$ & $1.7 \times 10^{-5} $ & $   1.2478$ \\ \hline
$[3/7;9]$ & $-0.1715697$ & $1.9 \times 10^{-5} $ & $   1.2464$ \\ \hline
$[3/8;6]$ & $-0.1715687$ & $2.4 \times 10^{-5} $ & $   1.2464$ \\ \hline
$[3/8;7]$ & $-0.1715696$ & $1.9 \times 10^{-5} $ & $   1.2474$ \\ \hline
$[3/8;8]$ & $-0.1715619$ & $6.4 \times 10^{-5} $ & $   1.2297$ \\ \hline
$[3/9;7]$ & $-0.1715680$ & $2.8 \times 10^{-5} $ & $   1.2436$ \\ \hline
$[4/4;3]$ & $-0.1715679$ & $2.9 \times 10^{-5} $ & $   1.2462$ \\ \hline
$[4/4;4]$ & $-0.1715608$ & $7.0 \times 10^{-5} $ & $   1.2446$ \\ \hline
$[4/4;5]$ & $-0.1715630$ & $5.7 \times 10^{-5} $ & $   1.2452$ \\ \hline
$[4/4;6]$ & $-0.1715669$ & $3.5 \times 10^{-5} $ & $   1.2464$ \\ \hline
$[4/5;3]$ & $-0.1715628$ & $5.9 \times 10^{-5} $ & $   1.2451$ \\ \hline
$[4/5;4]$ & $-0.1715632$ & $5.6 \times 10^{-5} $ & $   1.2452$ \\ \hline
$[4/5;5]$ & $-0.1715648$ & $4.7 \times 10^{-5} $ & $   1.2458$ \\ \hline
$[4/5;6]$ & $-0.1715670$ & $3.4 \times 10^{-5} $ & $   1.2465$ \\ \hline
$[4/5;7]$ & $-0.1715686$ & $2.5 \times 10^{-5} $ & $   1.2454$ \\ \hline
$[4/6;4]$ & $-0.1715663$ & $3.8 \times 10^{-5} $ & $   1.2462$ \\ \hline
$[4/6;5]$ & $-0.1715673$ & $3.2 \times 10^{-5} $ & $   1.2464$ \\ \hline
$[4/6;6]$ & $-0.1715689$ & $2.3 \times 10^{-5} $ & $   1.2465$ \\ \hline
$[4/6;7]$ & $-0.1715690$ & $2.3 \times 10^{-5} $ & $   1.2468$ \\ \hline
$[4/6;8]$ & $-0.1715635$ & $5.5 \times 10^{-5} $ & $   1.2360$ \\ \hline
$[4/7;5]$ & $-0.1715685$ & $2.5 \times 10^{-5} $ & $   1.2480$ \\ \hline
$[4/7;6]$ & $-0.1715689$ & $2.3 \times 10^{-5} $ & $   1.2466$ \\ \hline
$[4/7;7]$ & $-0.1715770$ & $2.4 \times 10^{-5} $ & $   1.2503$ \\ \hline
$[4/7;8]$ & $-0.1715900$ & $1.0 \times 10^{-4} $ & $   1.2273$ \\ \hline
$[4/8;6]$ & $-0.1715674$ & $3.2 \times 10^{-5} $ & $   1.2441$ \\ \hline 
\hline
\end{tabular}
\end{center}
\caption{Continuation of DA Table for $u_e$ Singularity.}
\label{table8c}
\end{table}

\begin{table}
\begin{center}
\begin{tabular}{|c|c|c|c|} \hline \hline
$[5/4;3]$ & $-0.1715611$ & $6.9 \times 10^{-5} $ & $   1.2446$ \\ \hline
$[5/4;4]$ & $-0.1715611$ & $6.9 \times 10^{-5} $ & $   1.2446$ \\ \hline
$[5/4;5]$ & $-0.1715654$ & $4.4 \times 10^{-5} $ & $   1.2459$ \\ \hline
$[5/4;6]$ & $-0.1715670$ & $3.4 \times 10^{-5} $ & $   1.2465$ \\ \hline
$[5/5;3]$ & $-0.1715607$ & $7.1 \times 10^{-5} $ & $   1.2445$ \\ \hline
$[5/5;4]$ & $-0.1715653$ & $4.4 \times 10^{-5} $ & $   1.2459$ \\ \hline
$[5/5;5]$ & $-0.1715674$ & $3.2 \times 10^{-5} $ & $   1.2462$ \\ \hline
$[5/5;6]$ & $-0.1715754$ & $1.4 \times 10^{-5} $ & $   1.2488$ \\ \hline
$[5/5;7]$ & $-0.1715682$ & $2.7 \times 10^{-5} $ & $   1.2456$ \\ \hline
$[5/6;4]$ & $-0.1715673$ & $3.3 \times 10^{-5} $ & $   1.2465$ \\ \hline
$[5/6;5]$ & $-0.1715662$ & $3.9 \times 10^{-5} $ & $   1.2490$ \\ \hline
$[5/6;6]$ & $-0.1715688$ & $2.4 \times 10^{-5} $ & $   1.2466$ \\ \hline
$[5/6;8]$ & $-0.1715835$ & $6.2 \times 10^{-5} $ & $   1.2433$ \\ \hline
$[5/7;5]$ & $-0.1715698$ & $1.8 \times 10^{-5} $ & $   1.2478$ \\ \hline
$[5/7;6]$ & $-0.1715665$ & $3.7 \times 10^{-5} $ & $   1.2427$ \\ \hline
$[5/7;7]$ & $-0.1715882$ & $8.9 \times 10^{-5} $ & $   1.2325$ \\ \hline
$[5/8;6]$ & $-0.1715723$ & $3.5 \times 10^{-6} $ & $   1.2495$ \\ \hline
$[6/4;2]$ & $-0.1715686$ & $2.5 \times 10^{-5} $ & $   1.2473$ \\ \hline
$[6/4;3]$ & $-0.1715610$ & $6.9 \times 10^{-5} $ & $   1.2446$ \\ \hline
$[6/4;4]$ & $-0.1715642$ & $5.1 \times 10^{-5} $ & $   1.2455$ \\ \hline
$[6/4;5]$ & $-0.1715667$ & $3.6 \times 10^{-5} $ & $   1.2463$ \\ \hline
$[6/4;6]$ & $-0.1715683$ & $2.7 \times 10^{-5} $ & $   1.2472$ \\ \hline
$[6/5;3]$ & $-0.1715655$ & $4.3 \times 10^{-5} $ & $   1.2459$ \\ \hline
$[6/5;4]$ & $-0.1715666$ & $3.6 \times 10^{-5} $ & $   1.2463$ \\ \hline
$[6/5;5]$ & $-0.1715697$ & $1.9 \times 10^{-5} $ & $   1.2464$ \\ \hline
$[6/5;6]$ & $-0.1715689$ & $2.3 \times 10^{-5} $ & $   1.2468$ \\ \hline
$[6/5;7]$ & $-0.1715662$ & $3.9 \times 10^{-5} $ & $   1.2420$ \\ \hline
$[6/6;4]$ & $-0.1715685$ & $2.5 \times 10^{-5} $ & $   1.2469$ \\ \hline
$[6/6;5]$ & $-0.1715689$ & $2.3 \times 10^{-5} $ & $   1.2467$ \\ \hline
$[6/6;6]$ & $-0.1715686$ & $2.5 \times 10^{-5} $ & $   1.2460$ \\ \hline
$[6/6;7]$ & $-0.1715819$ & $5.3 \times 10^{-5} $ & $   1.2461$ \\ \hline
$[6/7;5]$ & $-0.1715630$ & $5.8 \times 10^{-5} $ & $   1.2355$ \\ \hline
$[6/7;6]$ & $-0.1715746$ & $9.9 \times 10^{-6} $ & $   1.2505$ \\ \hline
$[7/4;2]$ & $-0.1715853$ & $7.2 \times 10^{-5} $ & $   1.2519$ \\ \hline
$[7/4;3]$ & $-0.1715648$ & $4.7 \times 10^{-5} $ & $   1.2457$ \\ \hline
$[7/4;4]$ & $-0.1715679$ & $2.9 \times 10^{-5} $ & $   1.2466$ \\ \hline
$[7/4;5]$ & $-0.1715679$ & $2.9 \times 10^{-5} $ & $   1.2466$ \\ \hline
$[7/4;6]$ & $-0.1715707$ & $1.3 \times 10^{-5} $ & $   1.2495$ \\ \hline
\hline
\end{tabular}
\end{center}
\caption{Continuation of DA Table for $u_e$ Singularity.}
\label{table8d}
\end{table}

\begin{table}
\begin{center}
\begin{tabular}{|c|c|c|c|} \hline \hline
$[7/5;3]$ & $-0.1715664$ & $3.8 \times 10^{-5} $ & $   1.2462$ \\ \hline
$[7/5;4]$ & $-0.1715679$ & $2.9 \times 10^{-5} $ & $   1.2466$ \\ \hline
$[7/5;5]$ & $-0.1715687$ & $2.4 \times 10^{-5} $ & $   1.2464$ \\ \hline
$[7/5;6]$ & $-0.1715678$ & $2.9 \times 10^{-5} $ & $   1.2446$ \\ \hline
$[7/5;7]$ & $-0.1715709$ & $1.2 \times 10^{-5} $ & $   1.2483$ \\ \hline
$[7/6;4]$ & $-0.1715691$ & $2.2 \times 10^{-5} $ & $   1.2471$ \\ \hline
$[7/6;5]$ & $-0.1715672$ & $3.3 \times 10^{-5} $ & $   1.2433$ \\ \hline
$[7/6;6]$ & $-0.1715652$ & $4.5 \times 10^{-5} $ & $   1.2388$ \\ \hline
$[7/7;5]$ & $-0.1715689$ & $2.3 \times 10^{-5} $ & $   1.2458$ \\ \hline
$[8/4;2]$ & $-0.1715791$ & $3.6 \times 10^{-5} $ & $   1.2505$ \\ \hline
$[8/4;3]$ & $-0.1715670$ & $3.4 \times 10^{-5} $ & $   1.2464$ \\ \hline
$[8/4;4]$ & $-0.1715679$ & $2.9 \times 10^{-5} $ & $   1.2466$ \\ \hline
$[8/4;5]$ & $-0.1715683$ & $2.7 \times 10^{-5} $ & $   1.2465$ \\ \hline
$[8/4;6]$ & $-0.1715609$ & $7.0 \times 10^{-5} $ & $   1.2283$ \\ \hline
$[8/5;3]$ & $-0.1715677$ & $3.0 \times 10^{-5} $ & $   1.2467$ \\ \hline
$[8/5;4]$ & $-0.1715682$ & $2.7 \times 10^{-5} $ & $   1.2465$ \\ \hline
$[8/5;5]$ & $-0.1715691$ & $2.2 \times 10^{-5} $ & $   1.2464$ \\ \hline
$[8/6;4]$ & $-0.1715700$ & $1.7 \times 10^{-5} $ & $   1.2476$ \\ \hline
$[8/6;5]$ & $-0.1715621$ & $6.3 \times 10^{-5} $ & $   1.2310$ \\ \hline
$[9/4;2]$ & $-0.1715693$ & $2.1 \times 10^{-5} $ & $   1.2472$ \\ \hline
$[9/4;3]$ & $-0.1715676$ & $3.1 \times 10^{-5} $ & $   1.2467$ \\ \hline
$[9/4;4]$ & $-0.1715684$ & $2.6 \times 10^{-5} $ & $   1.2466$ \\ \hline
$[9/4;5]$ & $-0.1715692$ & $2.2 \times 10^{-5} $ & $   1.2471$ \\ \hline
$[9/4;6]$ & $-0.1715572$ & $9.1 \times 10^{-5} $ & $   1.2152$ \\ \hline
$[9/5;3]$ & $-0.1715658$ & $4.1 \times 10^{-5} $ & $   1.2458$ \\ \hline
$[9/5;4]$ & $-0.1715691$ & $2.2 \times 10^{-5} $ & $   1.2469$ \\ \hline
$[9/5;5]$ & $-0.1715690$ & $2.3 \times 10^{-5} $ & $   1.2460$ \\ \hline
$[9/6;4]$ & $-0.1715709$ & $1.1 \times 10^{-5} $ & $   1.2483$ \\ \hline
$[10/4;2]$ & $-0.1715630$ & $5.7 \times 10^{-5} $ & $  1.2448$ \\ \hline
$[10/4;3]$ & $-0.1715686$ & $2.5 \times 10^{-5} $ & $  1.2470$ \\ \hline
$[10/4;4]$ & $-0.1715681$ & $2.8 \times 10^{-5} $ & $  1.2457$ \\ \hline
$[10/4;5]$ & $-0.1715643$ & $5.0 \times 10^{-5} $ & $  1.2385$ \\ \hline
$[10/5;3]$ & $-0.1715698$ & $1.8 \times 10^{-5} $ & $  1.2476$ \\ \hline
$[11/4;2]$ & $-0.1715688$ & $2.4 \times 10^{-5} $ & $  1.2471$ \\ \hline
$[11/4;3]$ & $-0.1715692$ & $2.2 \times 10^{-5} $ & $  1.2473$ \\ \hline
$[11/4;4]$ & $-0.1715669$ & $3.5 \times 10^{-5} $ & $  1.2433$ \\ \hline
$[11/5;3]$ & $-0.1715707$ & $1.2 \times 10^{-5} $ & $  1.2481$ \\ \hline
\hline
\end{tabular}
\end{center}
\caption{Continuation of DA Table for $u_e$ Singularity.}
\label{table8e}
\end{table}

\vfill
\eject

\begin{center}
{\bf Figure Caption}
\end{center}

 Fig. 1. \ Assignment of couplings $K_i$, $i=1,2,3,4$ to links (bonds) of the
checkerboard lattice. 

 Fig. 2. \  (a) Phases and associated boundaries in the complex $u$ plane 
for the Ising model on the square lattice with $h = \pm i\pi/2$.  
The boundaries are given by eqs. (\ref{ucircle}) and (\ref{usegment}) in the 
text.  In particular, the line segment extends from $u_e$ as given in eq. 
(\ref{ue}) off the figure to the left, ending at 
$1/u_e = -(3+2\sqrt{2}) \simeq -5.828$. FM and AFM refer to
phases in which $M \ne 0$, $M_{st}=0$ and $M=0$, $M_{st} \ne 0$, respectively.
(b) Complex-temperature phase diagram in the $v$ plane. 

Fig. 3.  \ Path of $\kappa^2$ corresponding to $u=\rho e^{i\theta}$ for
$\theta=\pi/6$, as $\rho$ varies from 0 to 30. Horizontal and vertical axes are
the $Re(\kappa^2)$ and $Im(\kappa^2)$ axes.  The image of the singular curve 
(\ref{ucircle}) and line segment (\ref{usegment}) is the dark line from 
$\kappa^2=1$ to $\kappa^2=\infty$. 

\vfill
\eject

\end{document}